\documentclass[
 reprint,
 showpacs,
 amsmath,amssymb,
 aps,
 prd,
 floatfix,
]{revtex4-1}

\usepackage{color}
\usepackage{graphicx}
\usepackage{lineno}
\usepackage{dcolumn}
\usepackage{bm}
\usepackage{subfigure}
\usepackage{multirow}

\newcommand{\BaBarYear}       {12}
\newcommand{\BaBarNumber}     {032}
\newcommand{\SLACPubNumber} {15374}
\newcommand{\BaBarType}      {PUB}  

\input{babarsym}

\long\def\inst#1{\par\nobreak\kern 4pt\nobreak
   {\it #1}\par\vskip 10pt plus 3pt minus 3pt}

\begin{document}

\preprint{\babar-PUB-\BaBarYear/\BaBarNumber} 
\preprint{SLAC-PUB-\SLACPubNumber} 

\begin{flushleft}
\babar-\BaBarType-\BaBarYear/\BaBarNumber \\
SLAC-PUB-\SLACPubNumber \\
arXiv:1304.5009\\
\vspace{0.4in}
\end{flushleft}

\title{\boldmath{Measurement of the $D^*(2010)^{+}$ Natural Line Width and the $D^*(2010)^{+} - D^0$ Mass Difference}}

%
\author{J.~P.~Lees}
\author{V.~Poireau}
\author{V.~Tisserand}
\affiliation{Laboratoire d'Annecy-le-Vieux de Physique des Particules (LAPP), Universit\'e de Savoie, CNRS/IN2P3,  F-74941 Annecy-Le-Vieux, France}
\author{E.~Grauges}
\affiliation{Universitat de Barcelona, Facultat de Fisica, Departament ECM, E-08028 Barcelona, Spain }
\author{A.~Palano$^{ab}$ }
\affiliation{INFN Sezione di Bari$^{a}$; Dipartimento di Fisica, Universit\`a di Bari$^{b}$, I-70126 Bari, Italy }
\author{G.~Eigen}
\author{B.~Stugu}
\affiliation{University of Bergen, Institute of Physics, N-5007 Bergen, Norway }
\author{D.~N.~Brown}
\author{L.~T.~Kerth}
\author{Yu.~G.~Kolomensky}
\author{G.~Lynch}
\affiliation{Lawrence Berkeley National Laboratory and University of California, Berkeley, California 94720, USA }
% ...
\collaboration{The \babar\ Collaboration}
\noaffiliation

\begin{abstract}
We measure the mass difference, $\Delta m_0$, between the $D^{*}(2010)^+$ and the $D^0$ and the natural line width, $\Gamma$, of the transition $D^{*}(2010)^+\to D^0 \pi^+$. The data were recorded with the \babar\ detector at center-of-mass energies at and near the $\Upsilon(4S)$ resonance, and correspond to an integrated luminosity of approximately $477 \invfb$. The $D^0$ is reconstructed in the decay modes $D^0 \to K^-\pi^+$ and $D^0 \to K^-\pi^+\pi^-\pi^+$. For the decay mode $D^0\to K^-\pi^+$ we obtain $\Gamma =  \left(83.4 \pm 1.7 \pm 1.5\right) \kev$ and $\Delta m_0 =  \left(145\,425.6 \pm 0.6 \pm 1.8\right) \kev$, where the quoted errors are statistical and systematic, respectively. For the $D^0\to K^-\pi^+\pi^-\pi^+$ mode we obtain $\Gamma = \left(83.2 \pm 1.5 \pm 2.6\right) \kev$ and $\Delta m_0 =  \left(145\,426.6 \pm 0.5 \pm 2.0\right) \kev$.  The combined measurements yield $\Gamma = \left(83.3 \pm 1.2 \pm 1.4\right) \kev$ and $\Delta m_0 = \left(145\,425.9 \pm 0.4 \pm 1.7\right) \kev$; the width is a factor of approximately 12 times more precise than the previous value, while the mass difference is a factor of approximately 6 times more precise.
\end{abstract}



\maketitle                                                                             

\setcounter{footnote}{0}


\section{Introduction}
\label{sec:Introduction}

The $D^{*}(2010)^{+}$ ($D^{*+}$) line width provides a window into a nonperturbative regime of strong physics where the charm quark is the heaviest meson constituent~\cite{Becirevic201394, actapolb.30.3849, Guetta2001134}. The line width provides an experimental check of models of the $D$ meson spectrum, and is related to the strong coupling of the $D^{*+}$ to the $D\pi$ system, $g_{D^* D \pi}$. In the heavy-quark limit, which is not necessarily a good approximation for the charm quark~\cite{PhysRevC.83.025205}, this coupling can be related to the universal coupling of heavy mesons to a pion, $\hat{g}$. There is no direct experimental window on the corresponding coupling in the $B$ system, $g_{B^*B\pi}$, since there is no phase space for the decay $B^* \to B\pi$.  However, the $D$ and $B$ systems can be related through $\hat{g}$, which allows the calculation of $g_{B^*B\pi}$. The $B^*B\pi$ coupling is needed for a model-independent extraction of $\left|V_{ub}\right|$~\cite{PhysRevD.49.2331,PhysRevLett.95.071802} and is presently one of the largest contributions to the theoretical uncertainty on $\left|V_{ub}\right|$~\cite{2009PhRvD79e4507B}.

We study the $D^{*+}\to D^0 \pi^+$ transition using the $D^0\to K^-\pi^+$ and $D^0\to K^-\pi^+\pi^-\pi^+$ decay modes to measure the values of the $D^{*+}$ line width, $\Gamma$, and the difference between the $D^{*+}$ and $D^0$ masses, $\Delta m_0$. The use of charge conjugate reactions is implied throughout this paper. The only prior measurement of the width is $\Gamma = \left(96 \pm 4 \pm 22\right) \kev$ by the CLEO collaboration where the uncertainties are statistical and systematic, respectively~\cite{PhysRevD.65.032003}. That measurement is based on a data sample corresponding to an integrated luminosity of $9 \invfb$ and reconstructed $D^0\to K^- \pi^+$ decays.  In the present analysis, we have a data sample that is approximately 50 times larger. This allows us to apply tight selection criteria to reduce background, and to investigate sources of systematic uncertainty with high precision.

The signal is described by a relativistic Breit-Wigner (RBW) function defined by

\begin{equation}
  \frac{d \Gamma(m)}{d m} =  \frac{m \Gamma_{D^*D \pi}\left(m\right) \, m_0 \Gamma}  {\left(m_0^2  - m^2\right)^2  + \left(m_0 \Gamma _{\text{Total}}(m) \right)^2},
  \label{eq:rbw}
\end{equation}
 
\noindent where $\Gamma_{D^*D \pi}$ is the partial width to $D^0\pi^+$, $m$ is the $D^0 \pi^+$ invariant mass, $m_0$ is the invariant mass at the pole, and $\Gamma_{\text{Total}}(m)$ is the total $D^{*+}$ decay width. The partial width is defined by 

\begin{equation}
  \Gamma_{D^*D \pi}(m) = \Gamma
  \left(\frac{\mathcal{F}^{\ell}_{D\pi}(p_0)}{\mathcal{F}^{\ell}_{D\pi}(p)}\right)^2\left(\frac{p}{p_0}\right)^{2\ell+1}\left(\frac{m_0}{m}\right),
  \label{eq:partialwidth}
\end{equation}

\noindent where $\mathcal{F}^{\ell = 1}_{D\pi}\left(p\right) = \sqrt{1+r^2 p^2}$ is the Blatt-Weisskopf form factor for a vector particle with radius parameter $r$ and daughter momentum $p$, and the subscript zero denotes a quantity measured at the pole~\cite{blatt, PhysRevD.5.624}. The value of the radius is unknown, but for the charm sector it is expected to be $\sim 1\gev^{-1}$~\cite{Albrecht1993435}.  We use the value $r = 1.6 \gev^{-1}$ from Ref.~\cite{Schwartz:2002hh} and vary this value as part of our investigation of systematic uncertainties.

The full width at half maximum (FWHM) of the RBW line shape ($\approx 100 \kev$) is much less than the FWHM of the almost Gaussian resolution function which describes more than 99\% of the signal ($\approx 300 \kev$).  Therefore, near the peak, the observed FWHM is dominated by the resolution function shape. However, the shapes of the resolution function and the RBW differ far away from the pole position. Starting $(1.5 - 2.0) \mev$ from the pole position, and continuing to $(5 - 10) \mev$ away (depending on the $ D^0 $ decay channel), the RBW tails are much larger. The signal rates in this region are strongly dominated by the intrinsic line width, not the resolution functions, and the integrated signals are larger than the integrated backgrounds. We use the very different resolution and RBW shapes, combined with the good signal-to-background rate far from the peak, to measure  $ \Gamma $ precisely.

The detailed presentation is organized as follows. Section~\ref{sec:detector} discusses the \babar\ detector and the data used in this analysis, and Section~\ref{sec:evtsel} describes the event selection. Section~\ref{sec:matmodel} discusses a correction to the detector material model and magnetic field map. Section~\ref{sec:fitstrategy} details the fit strategy, Section~\ref{sec:systematics} discusses and quantifies the sources of systematic uncertainty, and Section~\ref{sec:combmodes} describes how the results for the two $D^0$ decay modes are combined to obtain the final results. Finally, the results are summarized in Section~\ref{sec:conclusion}.


\section{\boldmath The \babar\ detector and data}
\label{sec:detector}
This analysis is based on a data sample corresponding to an integrated luminosity of approximately $477\invfb$ recorded at and $40 \mev$ below the $\Upsilon\left(4S\right)$ resonance by the \babar\ detector at the PEP-II asymmetric energy \epem\ collider~\cite{Lees2013203}. The \babar\ detector is described in detail elsewhere~\cite{ref:babar, ref:nim_update}, so we summarize only the relevant components below. Charged particles are measured with a combination of a 40-layer cylindrical drift chamber (DCH) and a 5-layer double-sided silicon vertex tracker (SVT), both operating within the $1.5$-T magnetic field of a superconducting solenoid. Information from a ring-imaging Cherenkov detector is combined with specific ionization $(dE/dx)$ measurements from the SVT and DCH to identify charged kaon and pion candidates. Electrons are identified, and photons measured, with a CsI(Tl) electromagnetic calorimeter. The return yoke of the superconducting coil is instrumented with tracking chambers for the identification of muons.

\section{Event selection}
\label{sec:evtsel}

We reconstruct continuum-produced $D^{*+}\rightarrow D^0 \pi_s^+$ decays in the two Cabibbo-favored channels $D^0 \to K^-\pi^+$ and $D^0\to K^-\pi^+\pi^-\pi^+$. The pion from the $D^{*+}$ decay is called the ``slow pion'' (denoted $\pi_s^+$) because of the limited phase space available. The mass difference of the reconstructed $D^{*+}$ and $D^{0}$ is denoted as $\Delta m$ (e.g. $m\left(K^-\pi^+\pi_s^+\right) -  m\left(K^-\pi^+\right)$ for the $D^0 \to K^-\pi^+$ channel). The resolution in $\Delta m$ is dominated by the resolution of the $\pi_s^+$ momentum, especially the uncertainty of its direction due to Coulomb multiple scattering. The selection criteria for the individual $D^0$ channels are detailed below; however, both modes have the same $D^{*+}$ requirements.  The selection criteria were chosen to enhance the signal-to-background ratio ($S/B$) to increase the sensitivity to the long RBW tails in the $\Delta m$ distribution; we have not optimized the criteria for statistical significance. Because this analysis depends on the RBW tails, we pay particular attention to how the selection criteria affect the tail regions.

The entire decay chain is fit using a kinematic fitter with geometric constraints at each vertex and the additional constraint that the $D^{*+}$ emerges from the luminous region, also referred to as the beam spot. The confidence level of the $\chi^2$ for this fit must must be greater than 0.1\%.  In addition, the confidence level  for the $\chi^2$ from fitting the $D^0$ daughter tracks to a common vertex must be at least 0.5\%. These confidence level selections reduce the set of final candidates by approximately 2.1\%. The beam spot constraint improves the $\Delta m$ resolution by a factor of 2.5, primarily because it constrains the direction of the $\pi^+_s$. If there is more than one $D^{*+}$ candidate in the event, we choose the one with the highest full decay chain confidence level. The reconstructed $D^0$ mass must be within the range $1.86 \gev$ to $1.87 \gev$. The mass difference between the $D^{*+}$ and $D^0$ is required to satisfy $\Delta m < 0.17 \gev$. A large amount of the combinatorial background is removed by requiring $p^*(D^{*+}) > 3.6 \gev$, where $p^*$ is the momentum measured in the \epem center-of-mass frame for the event.

\begin{figure}[h!]
\begin{center}
\includegraphics[scale=0.45]{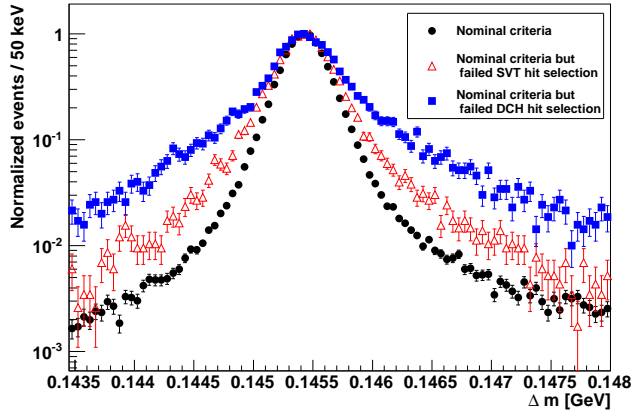}
\caption{(color online) Disjoint sets of $D^0 \to K^-\pi^+$ candidates illustrating the candidates that fail the tracking requirements have worse $\Delta m$ resolution. Each histogram is normalized to its peak. The events that populate the narrowest peak are the nominal $D^{*+}$ candidates that pass all selection criteria. The events that populate the intermediate and widest peaks pass all selection criteria except either the slow pion candidates or $D^0$ daughters fail the SVT requirements or fail the DCH requirements, respectively.}
\label{fig:tracking_cuts}
\end{center}
\end{figure}

\begin{figure}[h!]
\begin{center}
\includegraphics[scale=0.45]{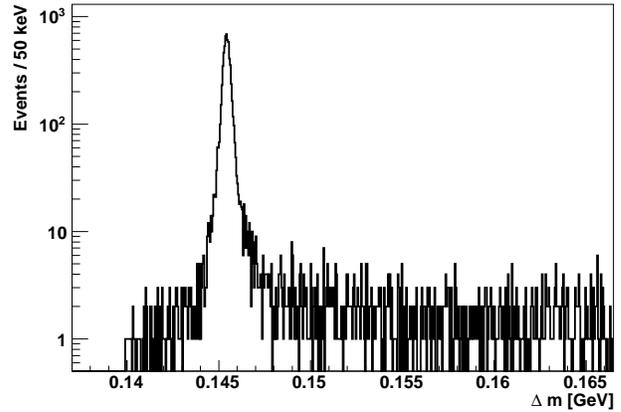}
\caption{Events with $D^{*+}$ candidates from $D^0\to K^- \pi^+$ that pass all selection criteria, but the slow pion candidate fails the $dE/dx$ requirement.}
\label{fig:pis_baddedx}
\end{center}
\end{figure}

To select well-measured slow pions we require that the $\pi_s^+$ tracks have at least $12$ measurements in the DCH and have at least 6 SVT measurements with at least 2 in the first three layers. For both $D^0\rightarrow K^- \pi^+$ and $D^0 \rightarrow K^- \pi^+\pi^-\pi^+$, we apply particle identification (PID) requirements to the $K$ and $\pi$ candidate tracks. To select candidates with better tracking resolution, and consequently improve the resolution of the reconstructed masses, we require that $D^0$ daughter tracks have at least 21 measurements in the DCH and satisfy the same SVT measurement requirements for the slow pion track. Figure~\ref{fig:tracking_cuts} illustrates the signal region distributions for three disjoint sets of $D^0 \to K^-\pi^+$ candidates: those passing all tracking requirements (narrowest peak), those otherwise passing all tracking requirements but failing the SVT hit requirements (intermediate peak), and those otherwise passing all tracking requirements but failing the requirement that both $D^0$ daughter tracks have at least 21 hits in the DCH and the $\pi_s^+$ track has at least 12 hits in the DCH (widest peak). The nominal sample (narrowest peak) has better resolution and S/B than candidates that fail the strict tracking requirements. We reduce backgrounds from other species of tracks in our slow pion sample by requiring that the $dE/dx$ values reported by the SVT and DCH be consistent with the pion hypothesis. Figure~\ref{fig:pis_baddedx} shows the $\Delta m$ distribution for candidates otherwise passing cuts, but in which the slow pion candidate fails either the SVT or DCH $dE/dx$ requirement. The $dE/dx$ selections remove protons from slow pion interactions in the beam pipe and detector material as well as electrons from the $D^{*0}$ decay chain discussed below. As shown in Fig.~\ref{fig:pis_baddedx}, while this requirement removes much more signal than background, the S/B ratio of the removed events is distinctly worse than that in the final sample.

\begin{figure}[h!]
\begin{center}
\includegraphics[scale=0.45]{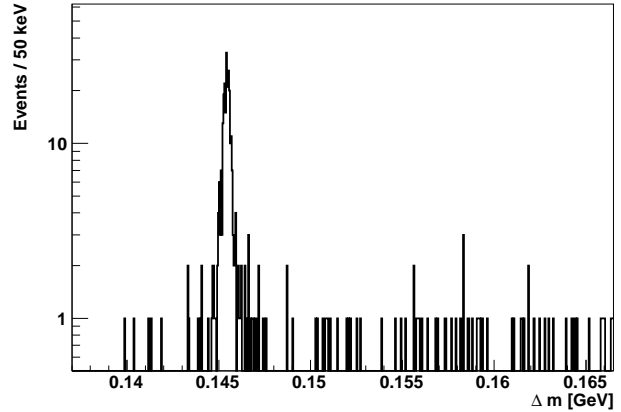}
\caption{Events with $D^{*+}$ candidates from $D^0\to K^- \pi^+$ that pass all selection criteria, but the slow pion candidate is identified by the algorithms as either a photon conversion in the detector material or a $\pi^0$ Dalitz decay.}
\label{fig:pis_badconv}
\end{center}
\end{figure}

The Dalitz decay $\pi^0\rightarrow \gamma e^+ e^-$ produces background where we misidentify an positron as a $\pi_s^+$.  We eliminate such candidates by reconstructing a candidate $e^+e^-$ pair and combining it with a $\gamma$. If the $e^+e^-$ vertex is within the SVT volume and the invariant mass is in the range $115 \mev< m\left(\gamma e^+ e^-\right) < 155 \mev$, then the event is rejected. Real photon conversions in the detector material are another source of background where electrons can be misidentified as slow pions. To identify such conversions we first create a candidate $e^+e^-$ pair using the slow pion candidate and an identified electron track from the same event and perform a least-squares fit with a geometric constraint. The event is rejected if the invariant mass of the putative pair is less than $60 \mev$ and the constrained vertex position is within the SVT tracking volume. Figure~\ref{fig:pis_badconv} shows the $\Delta m$ distribution for candidates otherwise passing cuts, but in which the slow pion candidate is identified as an electron using either of these $\pi^0$ conversion algorithms. As shown in Fig.~\ref{fig:pis_badconv}, only a small number of $D^{*+}$ candidates pass all other selection criteria but have a slow pion rejected by these algorithms. Again, the S/B ratio of this sample is distinctly worse than that of the final sample.

We identified additional criteria to remove candidates in kinematic regions where the Monte Carlo (MC) simulation poorly models the data. The MC is a cocktail of $q\bar{q}$ and $\ell^+ \ell^-$ sources where $q = u, d, s, c, b$ and $\ell = e, \mu, \tau$. The simulation does not accurately replicate the momentum distributions observed in data at very high and low $D^{*+}$ momentum values, so we require that $3.6 \gev < p^*(D^{*+}) < 4.3 \gev$ and that the laboratory momentum of the slow pion be at least $150 \mev$. In an independent sample of $K_{S}^{0}\to \pi^- \pi^+$ decays, the reconstructed $K_S^0$ mass is observed to vary as a function of the polar angle $\theta$ of the $K_S^{0}$ momentum measured in the laboratory frame with respect to the electron beam axis. We define the acceptance angle to reject events where any of the daughter tracks of the $D^{*+}$ has $\cos \theta \ge 0.89$ to exclude the very-forward region of the detector. This criterion reduces the final data samples by approximately 10\%. 

The background level in the $D^0 \rightarrow K^-\pi^+\pi^-\pi^+$ mode is much higher than that in $D^0 \rightarrow K^-\pi^+$, and so we require $D^0$ daughter charged tracks to satisfy stricter PID requirements. The higher background arises because the $D^0$ mass is on the tail of the two-body $K^-\pi^+$ invariant mass distribution expected in a longitudinal phase space model, however it is near the peak of the 4-body $K^-\pi^+\pi^-\pi^+$ invariant mass distribution~\cite{feynman1972}. In addition, there is more random combinatorial background in the 4-track $D^0 \to K^- \pi^+\pi^-\pi^+$ mode than in the 2-track $D^0 \to K^- \pi^+$ mode.

The initial fit to the $D^0 \to K^-\pi^+\pi^-\pi^+$ validation signal MC sample had a bias in the measured value of the $D^{*+}$ width. An extensive comparison revealed that the bias originated from regions of phase space that the MC generator populated more frequently than the data.  Evidently, there are amplitudes that suppress these structures in the data, that are neither known nor included in the MC generator. We avoid the regions where the MC disagrees with the data by rejecting a candidate if either $\left(m^2\left(\pi^+ \pi^+\right) < -1.17 \, m^2\left(\pi^- \pi^+\right) + 0.46 \gev^2\right)$ or $\left( m^2\left(\pi^-\pi^+\right)< 0.35 \gev^2\right.$ and $\left.m^2\left(K^-\pi^+\right) < 0.6 \gev^2 \right)$. This veto is applied for each $\pi^+$ daughter of the $D^0$ candidate. Including or excluding these events has no noticeable effect on the central values of the parameters from the data. These vetoes reduce the final candidates by approximately 20\%.

\begin{figure*}[!h]
\begin{center}
\subfigure{\includegraphics[scale=0.35]{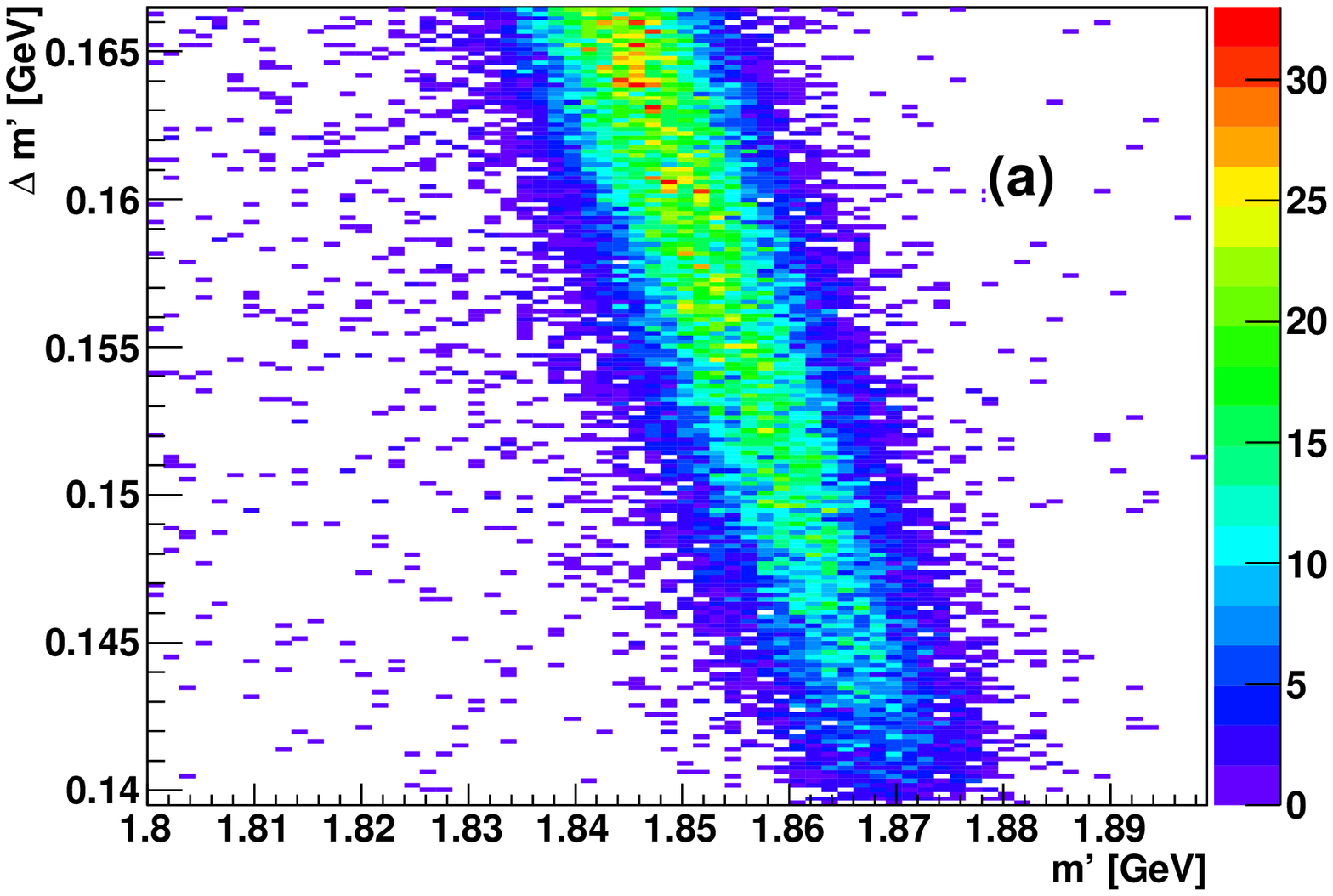}\label{fig:mdmprime_corr}}
\subfigure{\includegraphics[scale=0.35]{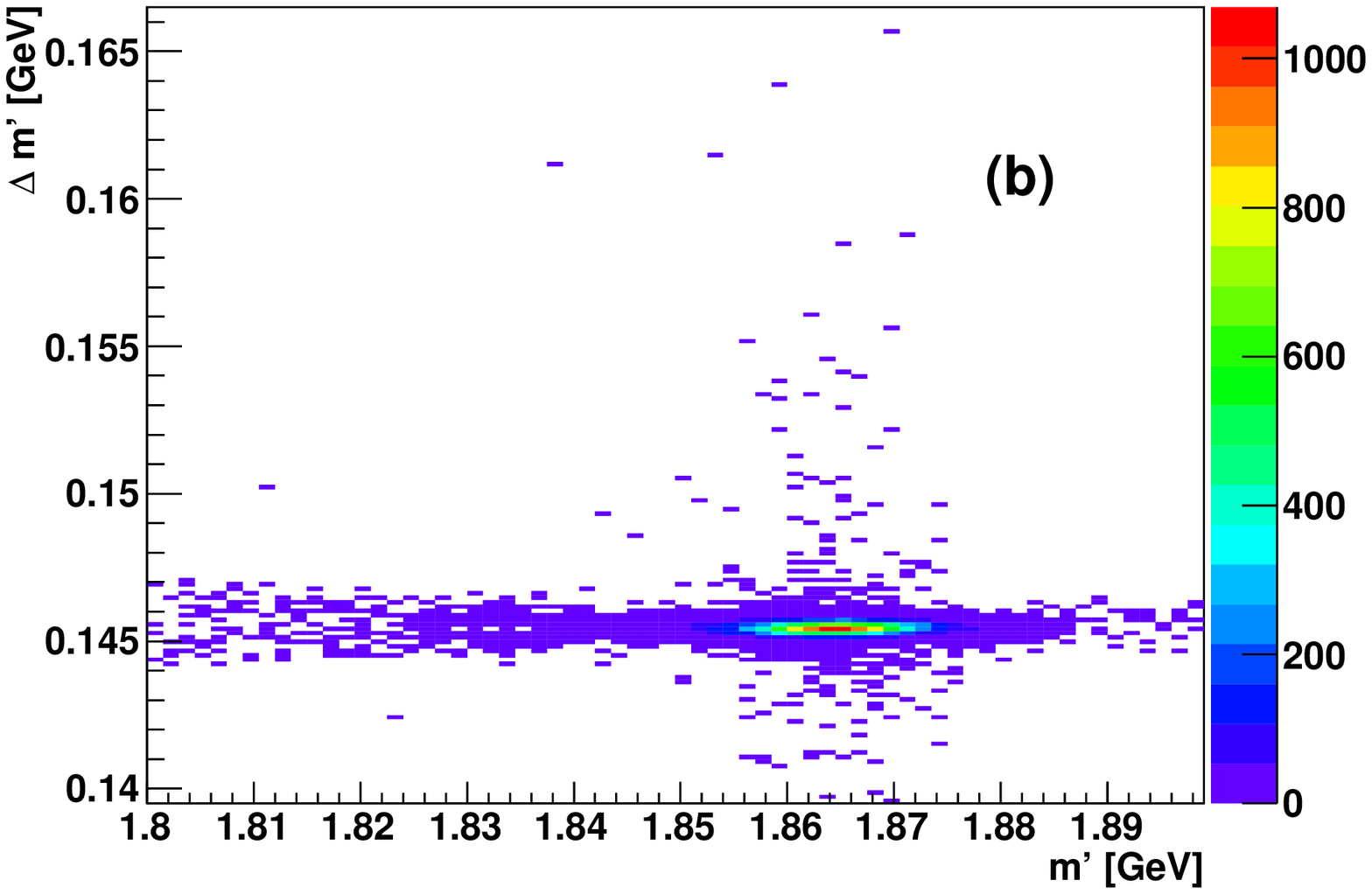}\label{fig:mdmprime_swap}} \\
\subfigure{\includegraphics[scale=0.35]{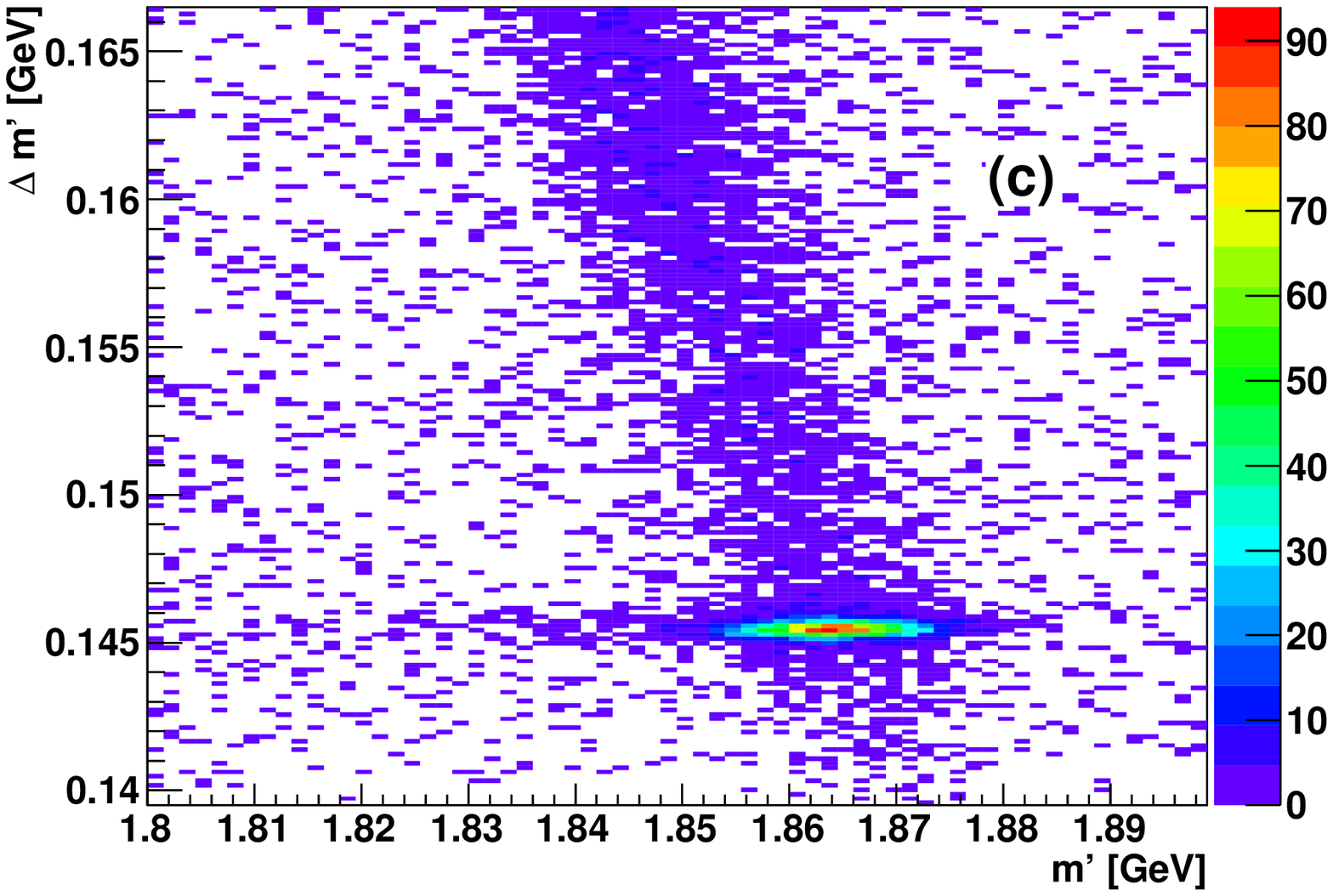}\label{fig:mdmprime_data}}
\end{center}
\caption{(color online) Illustrations of the $(m', \Delta m')$ system in (a) MC with the $D^{*+}$ correctly reconstructed, (b) MC with the slow pion and a $D^0$ daughter pion swapped during reconstruction, and (c) in data. The majority of correctly reconstructed decays are located outside of the shown $(m', \Delta m')$ range.}
\label{fig:mdmprime}
\end{figure*}

There is an additional source of background that must be taken into account for the $K^-\pi^+\pi^-\pi^+$ channel that is negligible for the $K^-\pi^+$ channel.  In a small fraction of events ($<1$\%) we mistakenly exchange the slow pion from $D^{*+}$ decay with one of the same-sign $D^0$ daughter pions. From the fits to the validation signal MC sample we find that this mistake would shift the reconstructed mass values and introduce a $\mathcal{O}(0.1 \kev)$ bias on the width. To veto these events we recalculate the invariant mass values after intentionally switching the same-sign pions, and create the variables $m' \equiv m\left(K^-\pi^+\pi^-\pi_s^+\right)$ and $\Delta m' \equiv m\left(K^-\pi^+\pi^-\pi^+\pi_s^+\right) -  m\left(K^-\pi^+\pi^-\pi_s^+\right)$. There are two pions from the $D^0$ decay with the same charge as the slow pion, so there are two values of $\Delta m'$ to consider.  In this procedure the correctly reconstructed events are moved away from the signal region, while events with this mis-reconstruction are shifted into the signal region. Figure~\ref{fig:mdmprime_corr} shows the $(m', \Delta m')$ distribution for MC events with correctly reconstructed $D^0$, where the majority of events are shifted past the bounds of the plot and only a small portion can be seen forming a diagonal band. The events with the slow pion and a $D^0$ daughter swapped are shown in Fig.~\ref{fig:mdmprime_swap} and form a clear signal. We reject events with $\Delta m' < 0.1665 \gev$.  Using fits to the validation signal MC sample, we find that this procedure removes approximately 80\% of the misreconstructed events and removes the bias reconstructed mass and the fitted value of the width. The $(m', \Delta m')$ distribution for data is shown in Fig.~\ref{fig:mdmprime_data}. Removing the $\Delta m'$ region reduces the final set of $D^0 \to K^-\pi^+\pi^-\pi^+$ candidates by approximately 2\%. The phase space distribution of events in MC and data differ slightly, so we expect differences in the efficiency of this procedure.

\section{Material modeling}
\label{sec:matmodel}

In the initial fits to data, we observed a very strong dependence of the RBW pole position on the slow pion momentum.  This dependence is not replicated in the MC, and originates in the magnetic field map and in the modeling of the material of the beam pipe and the SVT.  
Previous \babar\ analyses have observed the similar effects, for example the measurement of the $\Lambda_c^+$ mass~\cite{PhysRevD.72.052006}. In that analysis the material model of the SVT was altered in an attempt to correct for the energy loss and the under-represented small-angle multiple scattering (due to nuclear Coulomb scattering). However, the momentum dependence of the reconstructed $\Lambda_c^+$ mass could be removed only by adding an unphysical amount of material to the SVT. In this analysis we use a different approach to correct the observed momentum dependence and adjust track momenta after reconstruction.

\begin{figure}[h!]
\begin{center}
\includegraphics[scale=0.45]{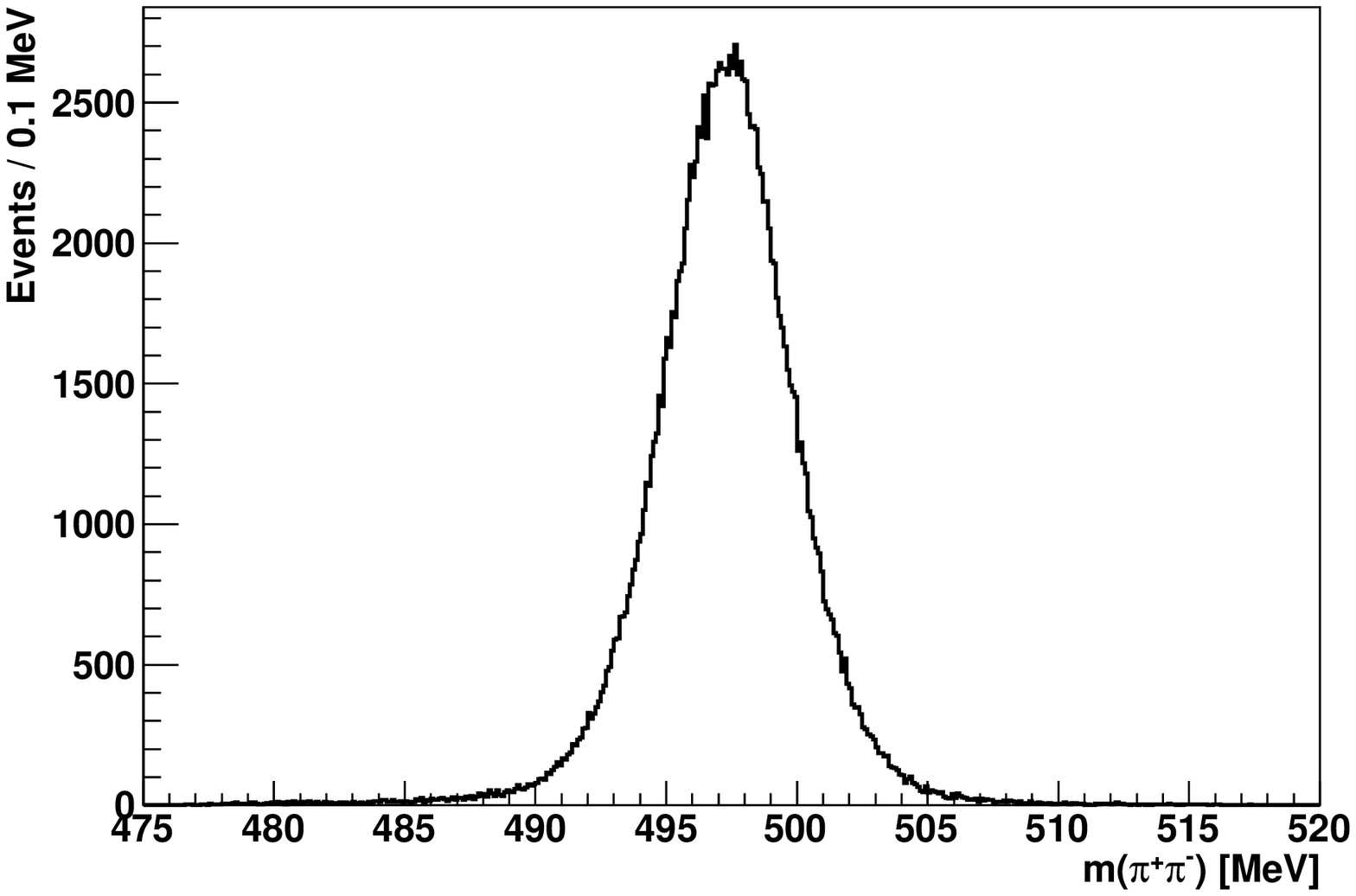}
\caption{Sample of $K_{S}^{0}\rightarrow \pi^+\pi^-$ candidates from $D^{*+}\to D^0 \pi_s^+ \to (K_{S}^{0} \pi^-\pi^+)\pi_s^+$ decay where the $K_{S}^{0}$ daughter pions satisfy the same tracking criteria as the slow pions of the $D^{*+}$ analysis.}
\label{fig:ksmass}
\end{center}
\end{figure}

\begin{figure}[h!]
\begin{center}
\includegraphics[scale=0.45]{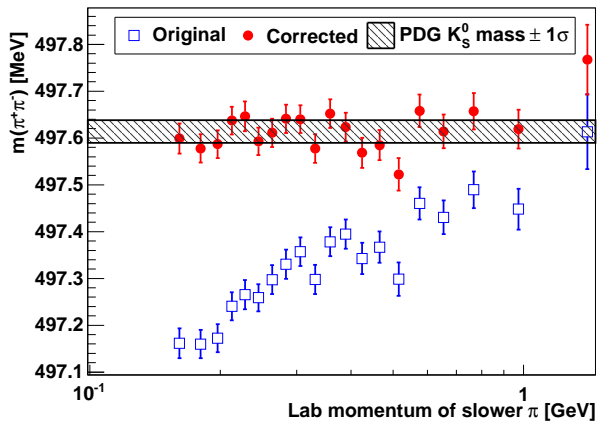}
\caption{(color online) Mass value of the $K_S^0$ obtained by fitting the invariant $\pi^+\pi^-$ mass distribution shown as a function of the slower pion laboratory momentum before (open squares) and after (closed circles) all energy-loss and momentum corrections have been applied. Note that the horizontal scale is logarithmic.}
\label{fig:ksmass_corr}
\end{center}
\end{figure}

We determine correction parameters using a sample of $K_{S}^{0}\rightarrow \pi^+\pi^-$ candidates from $D^{*+}\to D^0 \pi^+$ decay, where we reconstruct $D^0 \to K_{S}^{0} \pi^-\pi^+$. In this study we require that the $K_{S}^{0}$ daughter pions satisfy the same tracking criteria as the slow pions of the $D^{*+}$ analysis. The $K_{S}^0$ decay vertex is required to be inside the beam pipe and to be well-separated from the $D^{0}$ decay vertex. These selection criteria yield an extremely clean $K_{S}^0$ sample (approximately $160000$ candidates, $>99.5\%$ pure), which is shown in Fig.~\ref{fig:ksmass}. This sample is used to determine fractional corrections to the overall magnetic field and to the energy losses in the beam pipe ($E_{\text{loss}}^{\text{bmp}}$) and, separately, in the SVT ($E_{\text{loss}}^{\text{svt}}$). 
The points represented as open squares in Fig.~\ref{fig:ksmass_corr} show the strong dependence of the reconstructed $K_{S}^0$ mass on laboratory momentum. 
Adjusting only the estimated energy losses and detector material flattens the distribution, but there is still a remaining discrepancy. This discrepancy is shown by the open squares in Fig.~\ref{fig:ksmass_corr} at high momentum and indicates an overall momentum scale problem. These two effects lead us to consider corrections to the laboratory momentum and energy of an individual track of the form

\begin{align}
p&\rightarrow p\left(1+a\right) \notag \\
E&\rightarrow E+b_{{\text{bmp}}} E_{\text{loss}}^{\text{bmp}} +b_{{\text{svt}}} E_{\text{loss}}^{\text{svt}}
\label{eq:epcorr}
\end{align}

\noindent where the initial energy losses are determined by the Kalman filter based on the material model. To apply the correction to a pion track, the magnitude of the momentum is first recalculated using the pion mass hypothesis and the corrected energy as shown in Eq.~(\ref{eq:epcorr}) where the energy losses ($E_{\text{loss}}^{\text{bmp}}$ and $E_{\text{loss}}^{\text{SVT}}$) are taken from the original Kalman fit. Then, the momentum is scaled by the parameter $a$ shown in Eq.~(\ref{eq:epcorr}) and the energy of the particle is recalculated assuming the pion mass hypothesis. The order of these operations, correcting the energy first and then the momentum, or vice versa, has a negligibly small effect on the calculated corrected invariant mass. After both pion tracks' momenta are corrected the invariant mass is calculated. Then the sample is separated into 20 intervals of $K_S^0$ momentum. Figure~\ref{fig:ksmass_corr} shows $m(\pi^+\pi^-)$ as a function of the slower pion laboratory momentum and illustrates that the momentum dependence of the original sample (open squares) has been removed after all of the corrections (closed circles). We determine the best set of correction parameters to minimize the $\chi^2$ of the bin-by-bin mass difference between the $\pi^+\pi^-$ invariant mass and the current value of the $K^{0}_S$ mass ($m_{\text{PDG}}\left(K_{S}^0\right)\pm1\sigma_{\text{PDG}} = 497.614 \pm 0.024 \mev$)~\cite{ref:pdg2012}. 

To estimate the systematic uncertainty in values measured from corrected distributions, we find new parameter values by tuning the $\pi^+\pi^-$ invariant mass to the nominal $K^{0}_S$ mass shifted up and down by one standard deviation. 
These three sets of correction parameters are listed in Table~\ref{table:corr_params}. The resulting average reconstructed $K_{S}^{0}$ masses after correction are $497.589 \pm 0.007 \mev$, $497.612 \pm 0.007 \mev$, and $497.640 \pm 0.007 \mev$ for target masses $m_{\text{PDG}}(K_{S}^0)-1\sigma_{\text{PDG}}$, $m_{\text{PDG}}(K_{S}^0)$, and $m_{\text{PDG}}(K_{S}^0)+1\sigma_{\text{PDG}}$, respectively. As these average values are so well-separated we do not include additional systematic uncertainties from parameters that could describe the central value. The systematics studies of fit result variations in disjoint subsamples of laboratory momentum remain sensitive to our imperfect correction model.

\begin{table}
\begin{center}
\caption{Energy-loss and momentum correction parameters of  Eq.~(\ref{eq:epcorr}) which remove the momentum dependence of the reconstructed $K_{S}^{0}$ mass shown in Fig.~\ref{fig:ksmass_corr}. The nominal parameters shift the average reconstructed masses to be the PDG mean value, also shown in Fig.~\ref{fig:ksmass_corr}. To estimate the associated systematic uncertainty, the procedure was repeated to give average reconstructed $K_{S}^{0}$ masses $\pm 1 \sigma_{\text{PDG}}$ from the nominal value.}
\begin{tabular}{cc@{\hspace{4mm}}c@{\hspace{2mm}}c}
\hline \hline\\[-1.7ex]
& Nominal & \multicolumn{2}{c}{For systematics} \\[-1.7ex] \\ \hline \\[-1.7ex]
 & $m_{\text{PDG}}(K_{S}^0)$ & $m_{\text{PDG}}+1\sigma_{\text{PDG}}$ & $m_{\text{PDG}}-1\sigma_{\text{PDG}}$ \\[-1.7ex]\\ \hline \\[-1.7ex]
$a$ & 0.00030 & 0.00031 & 0.00019\\ 
$b_{\text{bmp}}$ & 0.0175 & 0.0517 & 0.0295\\ 
$b_{\text{svt}}$ & 0.0592 & 0.0590 & 0.0586 \\[-1.7ex]\\  \hline \hline
\end{tabular}
\label{table:corr_params}
\end{center}
\end{table}

The best-fit value of $a=0.00030$ corresponds to an increase of $4.5$ Gauss on the central magnetic field. This is larger than the nominal $2$ Gauss sensitivity of the magnetic field mapping~\cite{ref:babar}. However, the azimuthal dependence of $\Delta m_0$ (discussed in Sec.~\ref{sec:systematics}) indicates that the accuracy of the mapping may be less than originally thought.

The momentum dependence of $\Delta m_0$ in the initial results is ascribed to underestimating the $dE/dx$ loss in the beam pipe and SVT, which we correct using the factors $b_{\text{bmp}}$ ($1.8\%$) and $b_{\text{SVT}}$ ($5.9\%$). Typical $dE/dx$ losses for a minimum ionizing particle with laboratory momentum $2 \gev$ traversing the beam pipe and SVT at normal incidence are $4.4 \mev$. The corrections are most significant for low-momentum tracks. However, the corrections are applied to all $D^{*+}$ daughter tracks, not just to the slow pion. The momentum dependence is eliminated after the corrections are applied. All fits to data described in this analysis are performed using masses and $\Delta m$ values calculated using corrected 4-momenta.  The MC tracks are not corrected because the same field and material models used to propagate tracks are used during their reconstruction.


\section{\boldmath Fit method}
\label{sec:fitstrategy}

To measure $\Gamma$ we fit the $\Delta m$ peak (the signal) with a relativistic Breit-Wigner (RBW) function convolved with a resolution function based on a Geant4 MC simulation of the detector response~\cite{geant4}. As in previous analyses~\cite{PhysRevD.65.032003}, we approximate the total $D^{*+}$ decay width $\Gamma_{\text{Total}}(m) \approx \Gamma_{D^*D \pi}(m)$, ignoring the electromagnetic contribution from $D^{*+}\rightarrow D^+ \gamma$. This approximation has a negligible effect on the measured values as it appears only in the denominator of Eq.~(\ref{eq:rbw}). For the purpose of fitting the $\Delta m$ distribution we obtain $d \Gamma(\Delta m)/d \Delta m$ from Eqs. (\ref{eq:rbw}) and (\ref{eq:partialwidth}) by making the substitution $m = m(D^0) + \Delta m$, where $m(D^0)$ is the current average mass of the $D^0$ meson~\cite{ref:pdg2012}.

Our fitting procedure involves two steps. In the first step we model the resolution due to track reconstruction by fitting the $\Delta m$ distribution for correctly reconstructed MC events using a sum of three Gaussians and a function to describe the non-Gaussian component. The second step uses the resolution shape from the first step and convolves the Gaussian components with a relativistic Breit-Wigner of the form in Eq.~(\ref{eq:rbw}) to fit the $\Delta m$ distribution in data, and thus measure $\Gamma$ and $\Delta m_0$. We fit the $\Delta m$ distribution in data and MC from the kinematic threshold to $\Delta m = 0.1665 \gev$ using a binned maximum likelihood fit and an interval width of $50 \kev$. Detailed results of the fits are presented in the Appendix~\ref{app:fitresults}.

\subsection{\boldmath Modeling experimental resolution}
\label{sec:resfit}

\begin{figure}[!h]
\begin{center}
\subfigure{\includegraphics[scale=0.35]{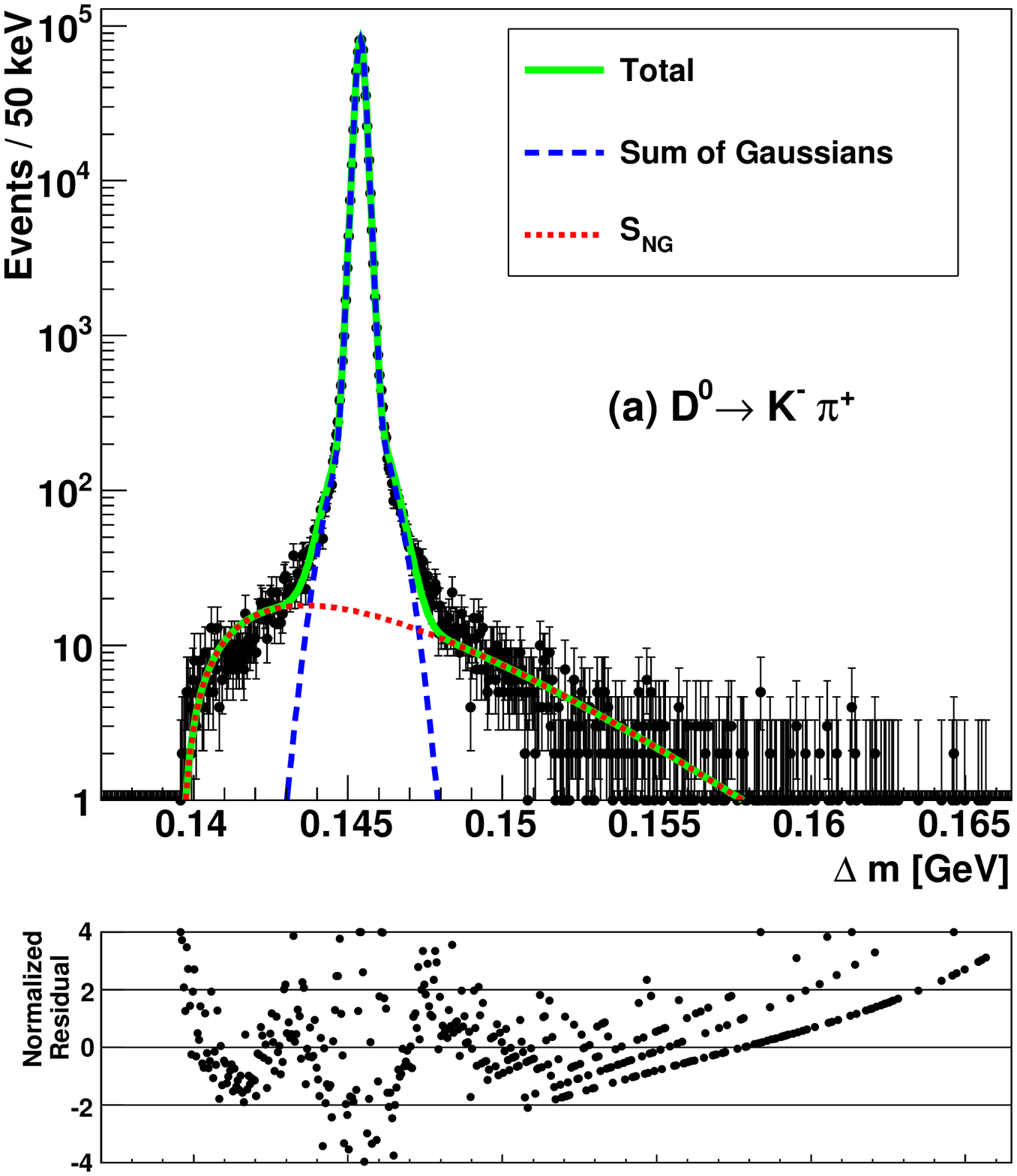}}
\subfigure{\includegraphics[scale=0.35]{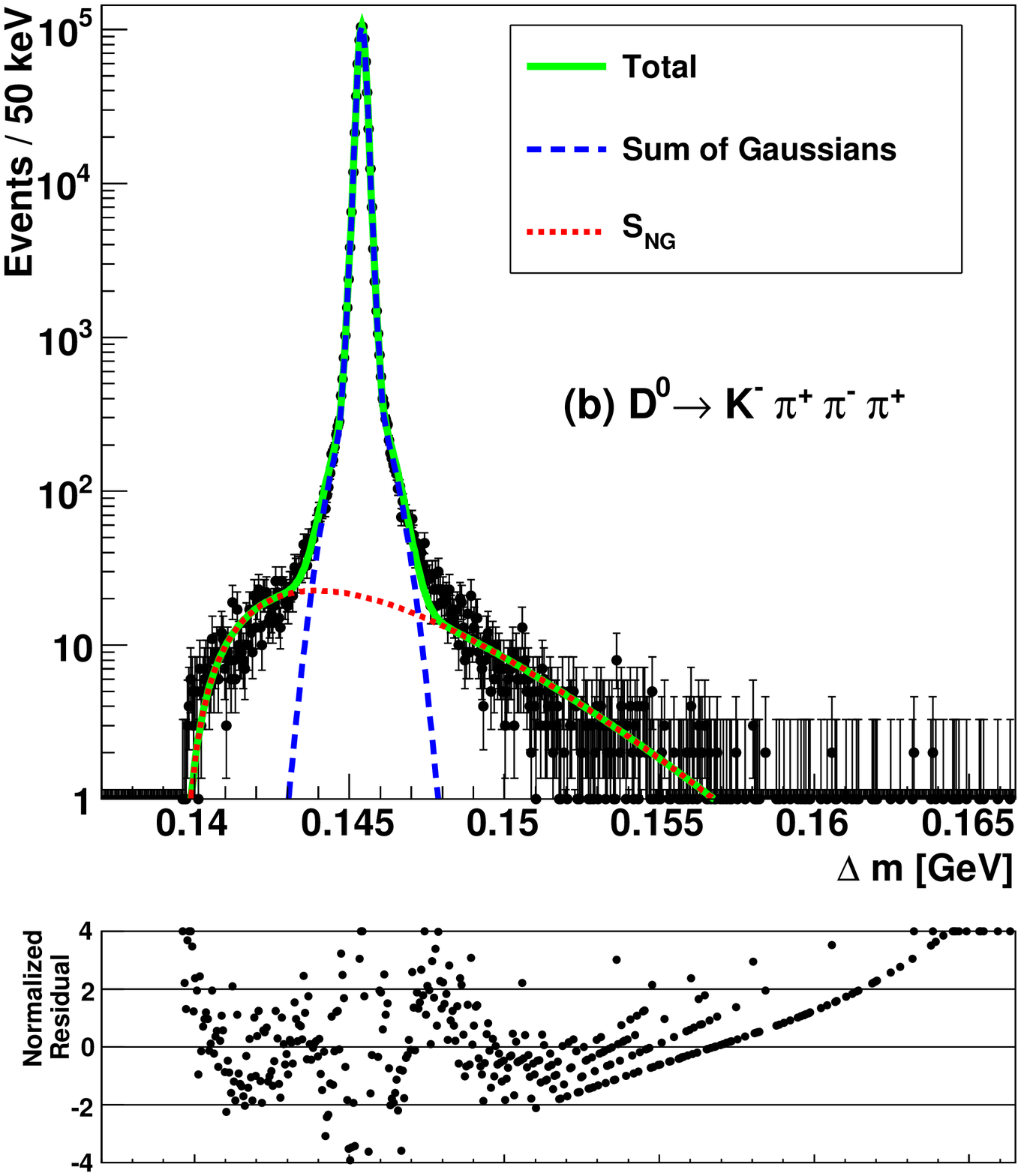}} \\
\end{center}
\caption{(color online) Binned maximum likelihood fit to the $\Delta m$ resolution distribution of MC samples for both $D^0$ decay modes. The interval size is $50 \kev$, and the high mass tails are dominated by low statistics. Normalized residuals are defined as $\left(N_{\text{observed}} - N_{\text{predicted}}\right)/\sqrt{N_{\text{predicted}}}$. The shapes in the distribution of the normalized residuals are from dominance by Poisson statistics. In the peak region the total PDF is visually indistinguishable from the Gaussian component of the resolution function.}
\label{fig:resfits}
\end{figure}

We generate samples of $D^{*+}$ decays with a line width of $0.1 \kev$, so that all of the observed spread is due to reconstruction effects.  The samples are approximately 5 times the size of the corresponding samples in data. The non-Gaussian tails of the distribution are from events in which the $\pi_s$ decays to a $\mu$ in flight and where coordinates from both the $\pi$ and $\mu$ segments are used in track reconstruction. Accounting for these non-Gaussian events greatly improves the quality of the fit to data near the $\Delta m$ peak.

We fit the $\Delta m$ distribution of the MC events with the function
\begin{align}
f_{NG} S_{NG}\left(\Delta m; q, \alpha \right)
 + (1 - f_{NG}) \left[f_1 G\left(\Delta m; \right.\right.&\left.\left.\mu_1, \sigma_1\right) \right. \nonumber \\ 
\left.+ f_2 G\left(\Delta m; \mu_2, \sigma_2\right)  + \left(1 - f_1 - f_2\right) G\left(\Delta m; \mu_3, \right.\right.&\left.\left.\sigma_3\right)\right]
\label{eq:respdf}
\end{align}
where the $G\left( \Delta m; \mu_i, \sigma_i \right)$ are Gaussian functions and $f_{NG}, f_1, f_2$ are the fractions allotted to the non-Gaussian component and the first and second Gaussian components, respectively.  The function describing the non-Gaussian component of the distribution is
\begin{equation}
S_{NG}\left(\Delta m; q, \alpha\right) = \Delta m \, u^q\,  e^{\alpha u},
\label{eq:resng}
\end{equation}
where $u \equiv \left(\Delta m/\Delta m_{\text{thres}}\right)^2 - 1$ and $\Delta m_{\text{thres}} = m_\pi$ is the kinematic threshold for the $D^{*+}\to D^0 \pi^+$ process. For $\Delta m < \Delta m_{\text{thres}}$, $S_{NG}$ is defined to be zero.

Figure~\ref{fig:resfits} shows the individual resolution function fits for the two $D^0$ decay modes.  Each plot shows the total resolution probability density function (PDF) as the solid curve, the sum of the Gaussian contributions is represented by the dashed curve, and the $S_{NG}$ function as a dotted curve describing the events in the tails.  The resolution functions should peak at the generated value, $\Delta m_0^{MC} = m(D^{*}(2010)^{+}) - m(D^0)$~\cite{ref:pdg2012}. However, the average value of the $\mu_i$ is slightly larger than the generated value of $\Delta m_0^{MC}$. The $S_{NG}$ function is excluded from this calculation as the peak position is not well defined and $S_{NG}$ describes less than 1\% of the signal. We take this reconstruction bias as an offset when measuring $\Delta m_0$ from data and denote this offset by $\delta m_0$. The $\delta m_0$ offset is $4.3 \kev$ and $2.8 \kev$ for the $D^0 \to K^-\pi^+$ and $D^0 \to K^-\pi^+\pi^-\pi^+$ modes, respectively.
As discussed in Sec.~\ref{sec:systematics}, although the values of $\delta m_0$ are larger than the final estimates of the systematic uncertainty for $\Delta m_0$, they are required for an unbiased result from fits to the validation signal MC samples. The systematic uncertainty associated with $\delta m_0$ is implicitly included when we vary the resolution shape, as discussed in Sec.~\ref{sec:systematics}. The parameter values, covariance matrix, and correlation matrix are present for each decay mode in the Appendix in Tables~\ref{tab:mcres}~-~\ref{tab:mcres_k3pi_corr}.

\subsection{\boldmath Fit Results}
\label{sec:datafit}

\begin{figure}[!h]
\centering
\subfigure{\includegraphics[scale=0.35]{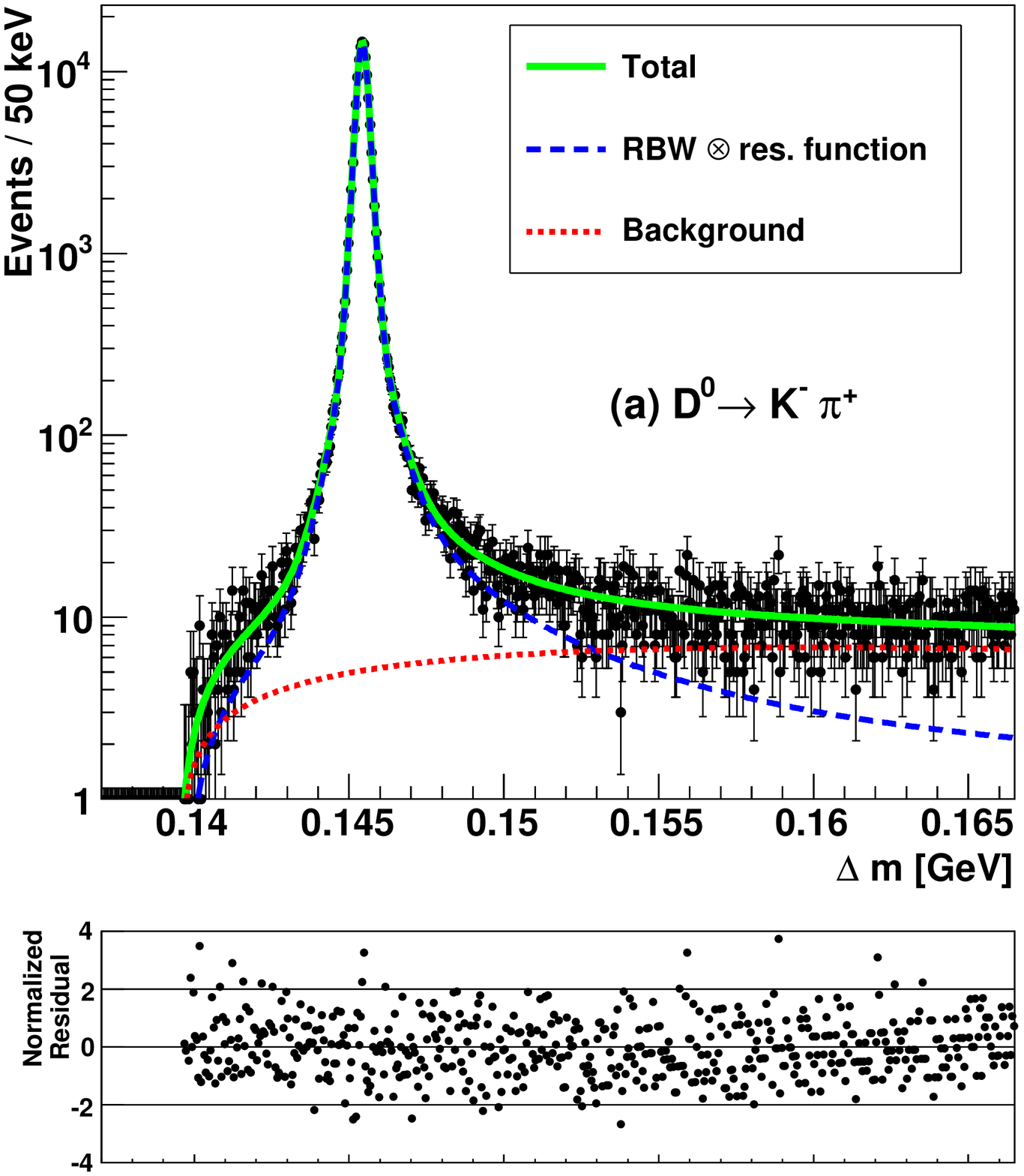}\label{fig:rdfits_kpi}}
\subfigure{\includegraphics[scale=0.35]{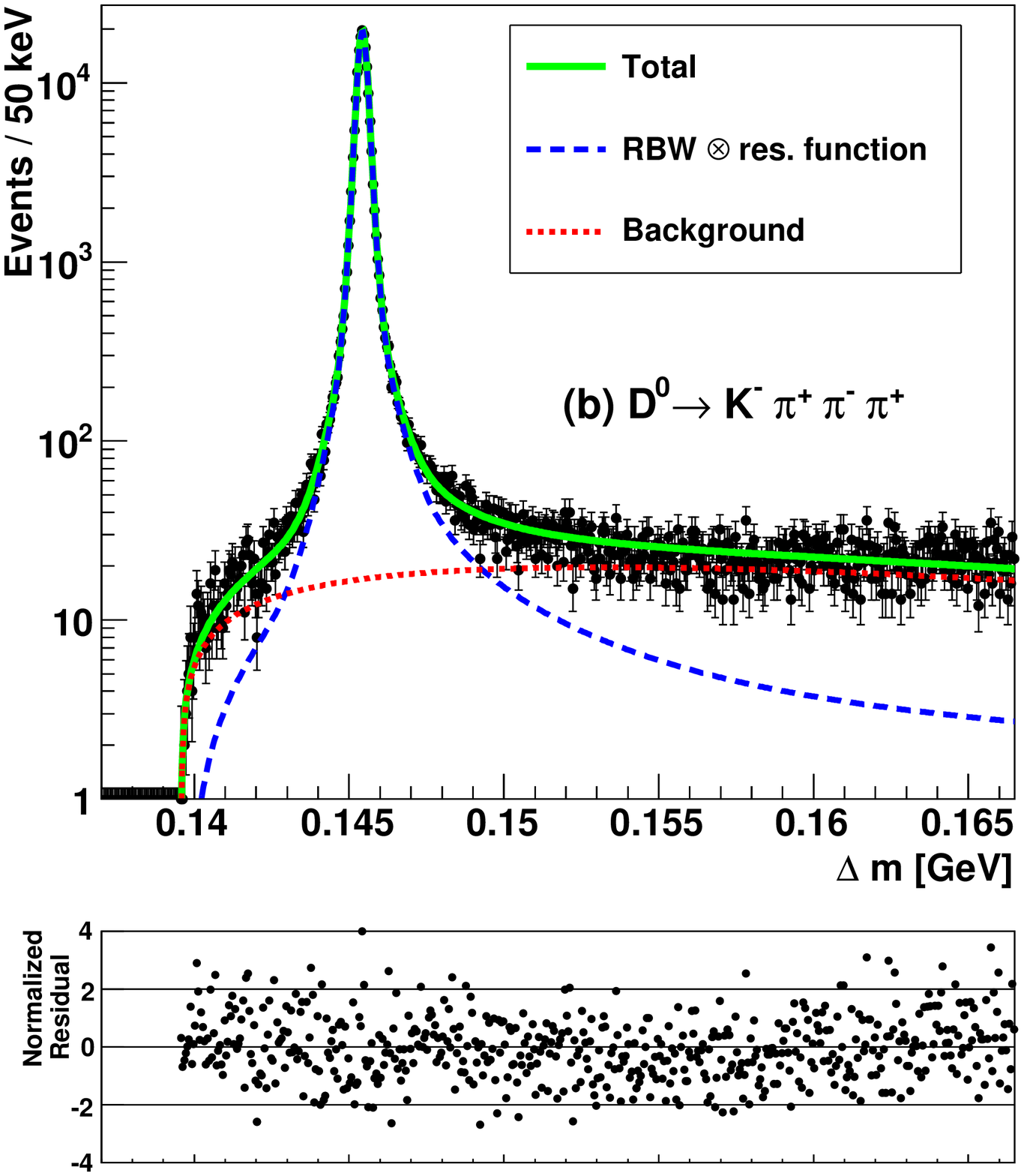}\label{fig:rdfits_k3pi}} \\
\caption{(color online) The results of the fits to data for each $D^0$ decay mode. The fitted parameter values are summarized in Table~\ref{table:kpik3pi_rd_summary}. The solid curve is the sum of the signal (dashed curve) and background (dotted curve) PDFs. The total PDF and signal component are visually indistinguishable in the peak region.}
\label{fig:rdfits}
\end{figure}

The parameters of the resolution function found in the previous step are used to create a convolved RBW PDF. 
In the fit to data, $S_{NG}$ has a fixed shape and relative fraction, and is not convolved with the RBW. 
The relative contribution of $S_{NG}$ is small and the results from the fits to the validation signal MC samples are unbiased without convolving this term. We fit the data using the function,

\begin{align}
\mathcal{P}(&\Delta m; \epsilon, \Gamma, \Delta m_0, c) = \nonumber \\
&f_{\mathcal{S}} \frac{\mathcal{S}(\Delta m; \epsilon, \Gamma, \Delta m_0)}{\int{\mathcal{S}(\Delta m)\, d \left(\Delta m\right)}}+(1-f_{\mathcal{S}}) \frac{\mathcal{B}(\Delta m; c)}{\int{\mathcal{B}(\Delta m)\, d \left(\Delta m\right)}}
\end{align}
where $f_{\mathcal{S}}$ is the fraction of signal events, $\mathcal{S}$ is the signal function
\begin{align}
\mathcal{S}(\Delta m) &= RBW \otimes  \nonumber \\
(1 - f_{NG}^{MC})&  \left[f_1^{MC} G\left(\Delta m; \mu_1^{MC} - \Delta m_0^{MC}, \sigma_1^{MC} \left(1+\epsilon\right)\right) \right. \nonumber \\
  \left. + f_2^{MC} \right.&\left.G\left(\Delta m; \mu_2^{MC} - \Delta m_0^{MC}, \sigma_2^{MC}\left(1+\epsilon\right)\right) \right. \nonumber \\
  \left. + \left(1 - \right.\right.&\left.\left.f_1^{MC} - f_2^{MC}\right) \right. \nonumber \\ 
\left. \right.&\left.G\left(\Delta m; \mu_3^{MC}- \Delta  m_0^{MC}, \sigma_3^{MC}\left(1+\epsilon\right)\right)\right] \nonumber \\
  + f_{NG}^{MC} &S_{NG}(\Delta m; q^{MC}, \alpha^{MC}),
 \label{eq:sigpdf}
\end{align}
and $\mathcal{B}$ is the background function
\begin{equation}
\mathcal{B}(\Delta m) = \Delta m \, \,  \sqrt{u} \, \, e^{c u},
 \label{eq:bkgpdf}
\end{equation}
\noindent where, again, $u \equiv \left(\Delta m/\Delta m_{\text{thres}}\right)^2-1$. The nominal RBW function has a pole position located at $ m = \Delta m_0 + m(D^0) $ and natural line width $ \Gamma $. The Gaussian resolution functions convolved with the RBW have centers offset from zero by small amounts determined from MC, $ \mu_i - \Delta m_0^{MC} $ (see Table~\ref{tab:mcres} in the Appendix). The widths determined from MC, $\sigma_i^{MC} $, are scaled by $ (1 + \epsilon ) $ where $ \epsilon $ is a common, empirically determined constant which accounts for possible differences between resolutions in data and simulation. As indicated in Eq.~(\ref{eq:sigpdf}), the parameters allowed to vary in the fit to data are the scale factor $(1+\epsilon)$, the width $\Gamma$, pole position $\Delta m_0$ and background shape parameter $c$. The validation of the fit procedure is discussed in Sec. \ref{sec:validation}.

Figure~\ref{fig:rdfits} shows the fits to data for both $D^0$ decay modes. The total PDF is shown as the solid curve, the convolved RBW-Gaussian signal as the dashed curve, and the threshold background as the dotted curve. The normalized residuals show the good agreement between the data and the model.  Table~\ref{table:kpik3pi_rd_summary} summarizes the results of the fits to data for the two modes. The covariance and correlation matrices for each mode are presented in Tables~\ref{tab:rd_kpi_cov}~-~\ref{tab:rd_k3pi_corr} in the Appendix. The tails of the RBW are much longer than the almost Gaussian resolution function. The resolution functions determined from the fits to MC drop by factors of more than 1000 near $\Delta m \approx 147 \mev$ with respect to the peak. At $\Delta m = 148 \mev$ the resolution functions have dropped by another factor of 10 and are dominated by the $S_{NG}$ component. The resolution functions used in fitting the data allow the triple-Gaussian part of the resolution function to scale by $(1+\epsilon)$, but the events observed above $148 \mev$ are predominantly signal events from the RBW tails and background. The signal from a zero-width RBW would approach 3 events per bin (see Fig.~\ref{fig:resfits}). The observed signal levels are of order 30 events per bin (see Fig.~\ref{fig:rdfits}). Table~\ref{table:kpik3pi_rd_summary} also shows the fitted $S/B$ at the peak and in the $\Delta m$ tail on the high side of the peak. The long non-Gaussian tail of the RBW is required for the model to fit the data so well.

As the observed FWHM values from the resolution functions are greater than the intrinsic line width, the observed widths of the central peaks determine the values of $\epsilon$. The scale factor, $(1+\epsilon)$, allows the resolution functions to expand as necessary to describe the distribution in real data. As one naively expects, the fitted values of the scale factor are strongly anti-correlated with the values for $\Gamma$ (the typical correlation coefficient is -0.85).

\begin{table}[!h]
\centering
\caption{Summary of the results from the fits to data for the $D^0\to K^-\pi^+$ and $D^0\to K^-\pi^+\pi^-\pi^+$ channels (statistical uncertainties only); $S/B$ is the ratio of the convolved signal PDF to the background PDF at the given value of $\Delta m$, and $\nu$ is the number of degrees of freedom.}
\begin{tabular}{c@{\hspace{4mm}}c@{\hspace{4mm}}c}
\hline \hline \\ [-1.7ex]
Parameter & $D^0\to K\pi$ & $D^0\to K\pi\pi\pi$ \\ \hline \\[-1.7ex]
Number of signal events & $138\,536 \pm 383$ & $174\,297\pm 434$ \\ 
$\Gamma \,(\kev)$ & $83.3 \pm 1.7$ & $83.2 \pm 1.5$ \\ 
scale factor, $(1+\epsilon)$ & $1.06 \pm 0.01$ & $1.08 \pm 0.01$ \\ 
$\Delta m_0\,(\kev)$ & $145\,425.6 \pm 0.6$ & $145\,426.6 \pm 0.5$ \\
background shape, $c$ &  $-1.97 \pm 0.28$ & $-2.82 \pm 0.13$ \\ \\[-1.3ex]
$S/B$ at peak & \multirow{2}{*}{$2700$} & \multirow{2}{*}{$1130$} \\ 
($\Delta m = 0.14542 \,(\gev)$) & & \\ \\ [-1.8ex]
$S/B$ at tail & \multirow{2}{*}{$0.8$} & \multirow{2}{*}{$0.3$} \\
($\Delta m = 0.1554\, (\gev)$) & & \\ \\ [-1.8ex]
$\chi^2/\nu$ & $574/535$ & $556/535$ \\[-1.7ex] \\ \hline \hline
\end{tabular}
\label{table:kpik3pi_rd_summary}
\end{table}


\section{\boldmath Systematic Uncertainties}
\label{sec:systematics}

\begin{table*}[!ht]
\centering
\caption{Summary of systematic uncertainties with correlation, $\rho$, between the $D^0\to K^-\pi^+$ and $D^0\to K^-\pi^+\pi^-\pi^+$ modes. The $K^-\pi^+$ and $K^-\pi^+\pi^-\pi^+$ invariant masses are denoted by $m\left(D^0_{\text{reco}}\right)$. The methods used to calculate or define the correlations are described in Sec.~\ref{sec:detcorr}. The total systematic uncertainties are calculated according to the procedure defined in Sec.~\ref{sec:combmodes}.}
\begin{tabular}{c@{\hspace{8mm}}ccc@{\hspace{8mm}}ccc}
\hline \hline  \\[-1.7ex] 
\multirow{2}{*}{Source} & \multicolumn{2}{c}{$\sigma_{sys} \left(\Gamma\right) \,[\kev]$} & \multirow{2}{*}{$\rho$} & \multicolumn{2}{c}{$\sigma_{sys} \left(\Delta m_0\right) [\kev]$}  & \multirow{2}{*}{$\rho$}\\  \\[-1.7ex] 
& $K\pi$ & $K\pi\pi\pi$ & & $K\pi$ & $K\pi\pi\pi$ & \\ \hline \\[-1.7ex]
Disjoint $p$ variation & 0.88 & 0.98 & \phantom{-}0.47 & 0.16 & 0.11 & 0.28 \\ 
Disjoint $m\left(D^0_{\text{reco}}\right)$ variation & 0.00 & 1.53 & \phantom{-}0.56 & 0.00 & 0.00 & 0.22 \\ 
Disjoint azimuthal variation & 0.62 & 0.92 & -0.04 & 1.50 & 1.68 & 0.84 \\ 
Magnetic field and material model & 0.29 & 0.18 & \phantom{-}0.98 & 0.75 & 0.81 & 0.99 \\ 
Blatt-Weisskopf radius & 0.04 & 0.04 & \phantom{-}0.99 & 0.00 & 0.00 & 1.00 \\ 
Variation of resolution shape parameters & 0.41 & 0.37 & \phantom{-}0.00 & 0.17 & 0.16 & 0.00  \\ 
$\Delta m$ fit range & 0.83 & 0.38 & -0.42 &  0.08 & 0.04 & 0.35  \\ 
Background shape near threshold & 0.10  & 0.33  & \phantom{-}1.00 & 0.00 & 0.00 & 0.00  \\ 
Interval width for fit & 0.00 & 0.05 & \phantom{-}0.99 & 0.00 & 0.00 & 0.00 \\ 
Bias from validation & 0.00 & 1.50 & \phantom{-}0.00 & 0.00 & 0.00 & 0.00 \\
Radiative effects & 0.25 & 0.11 & \phantom{-}0.00 & 0.00 & 0.00 & 0.00 \\ \hline \\[-1.7ex]
Total & 1.5 & 2.6 & & 1.7 & 1.9 & \\[-1.7ex]\\ \hline \hline
\end{tabular}
\label{table:syswithcorr}
\end{table*}

We estimate systematic uncertainties associated with instrumental effects by looking for large variations of results in disjoint subsets. The systematic uncertainties associated with our fit procedure are estimated using a variety of techniques. These methods are summarized in the following paragraphs and then discussed in detail.

To estimate systematic uncertainties from instrumental effects, we divide the data into disjoint subsets corresponding to intervals of laboratory momentum, $p$, of the $D^{*+}$, azimuthal angle, $\phi$, of the $D^{*+}$ in the laboratory frame, and reconstructed $D^0$ mass. In each of these variables we search for variations greater than those expected from statistical fluctuations.

After the corrections to the material model and magnetic field, the laboratory momentum dependence of the RBW pole position is all but eliminated. We find that $\Gamma$ does not display an azimuthal dependence, however $\Delta m_0$ does. Neither $\Gamma$ nor $\Delta m_0$ displays a clear systematic shape with reconstructed $D^0$ mass.

The uncertainties associated with the various parts of the fit procedure are investigated in detail. We vary the parameters of the resolution function in Eq.~(\ref{eq:respdf}) according to the covariance matrix reported by the fit to estimate systematic uncertainty of the resolution shape. Changing the end point for the fit estimates a systematic uncertainty associated with the shape of the background function. We also change the background shape near threshold. To estimate the uncertainty in the Blatt-Weisskopf radius we model the $D^{*+}$ as a point-like particle. We fit MC validation samples to estimate systematic uncertainties associated with possible biases. Finally, we estimate possible systematic uncertainties due to radiative effects. All of these uncertainties are estimated independently for the $D^0\rightarrow K^-\pi^+$ and $D^0\rightarrow K^-\pi^+\pi^-\pi^+$ modes, and are summarized in Table~\ref{table:syswithcorr}.

\begin{figure*}[p]
   \begin{center}
   \subfigure{\includegraphics[scale=0.43]{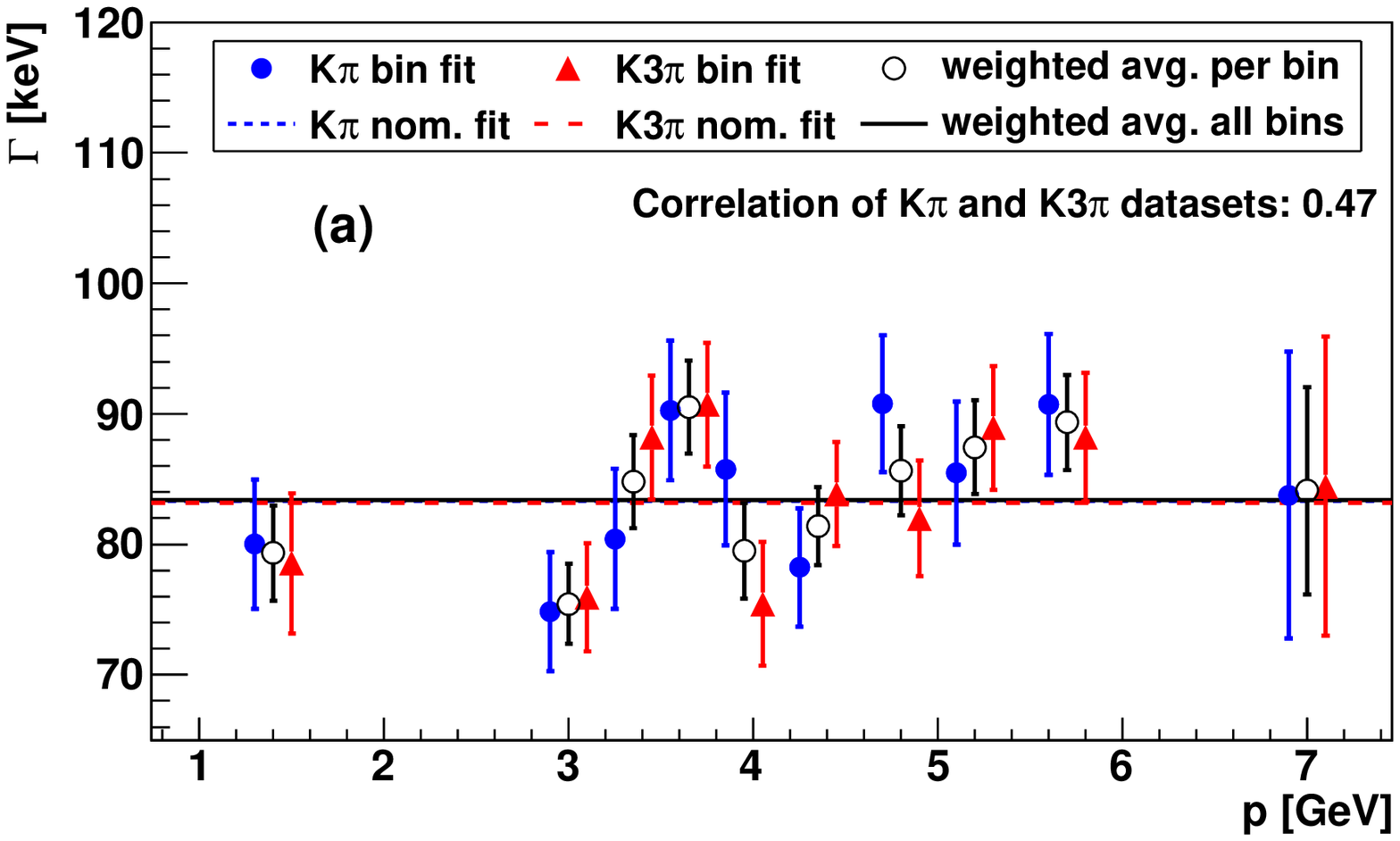} \label{fig:plabwidth}} 
   \subfigure{\includegraphics[scale=0.43]{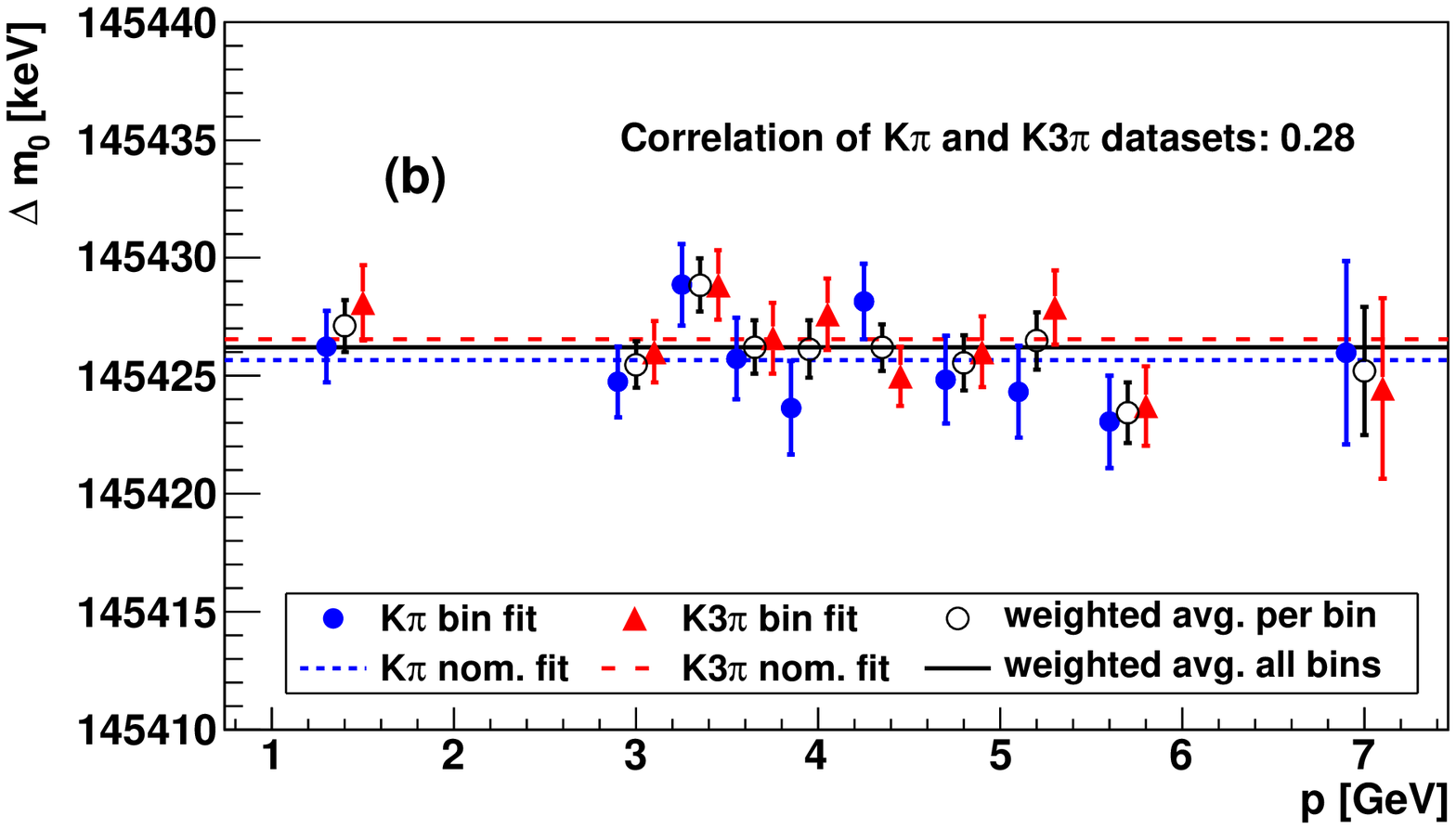} \label{fig:plabrbw}} \\
   \subfigure{\includegraphics[scale=0.43]{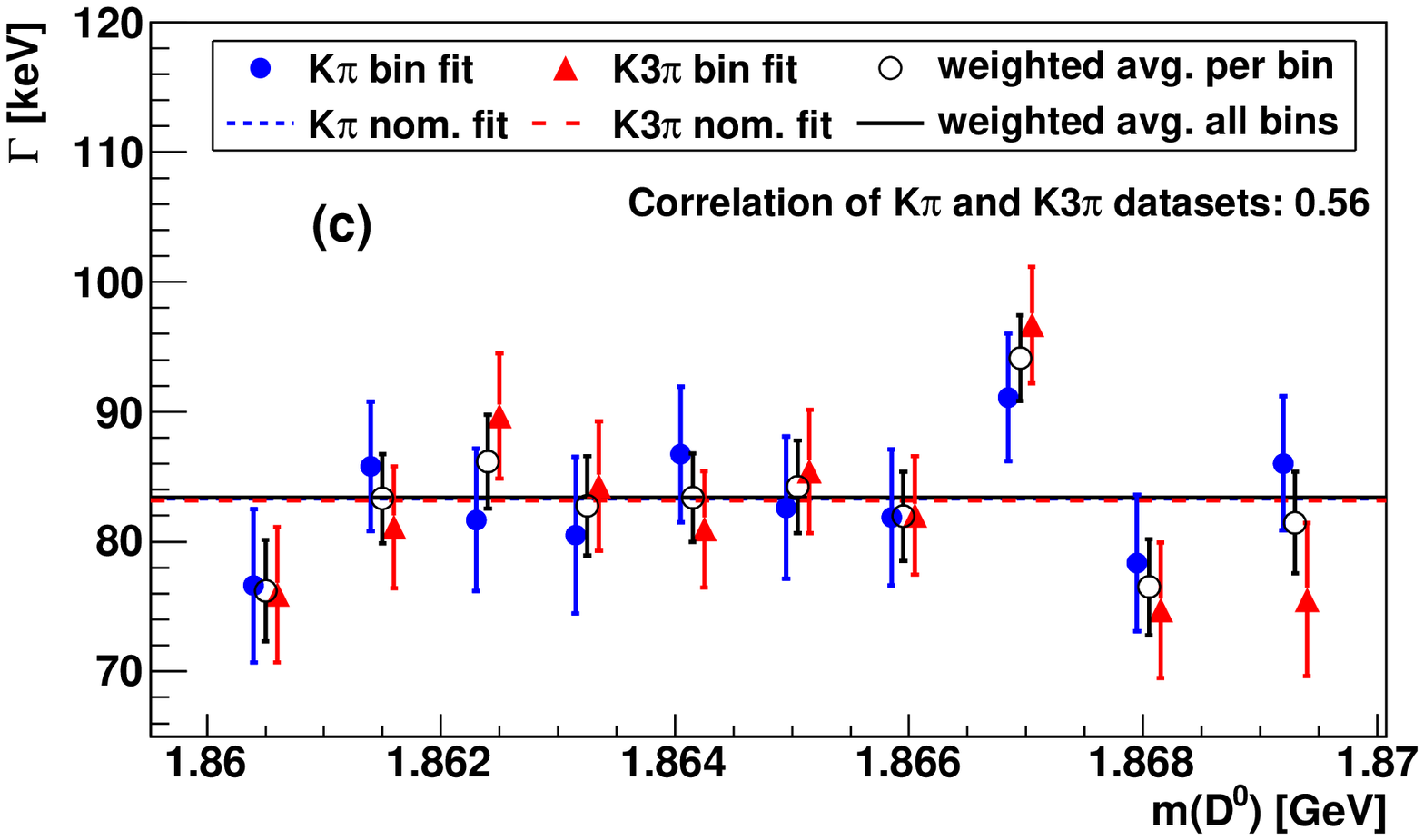} \label{fig:mslicewidth}}
   \subfigure{\includegraphics[scale=0.43]{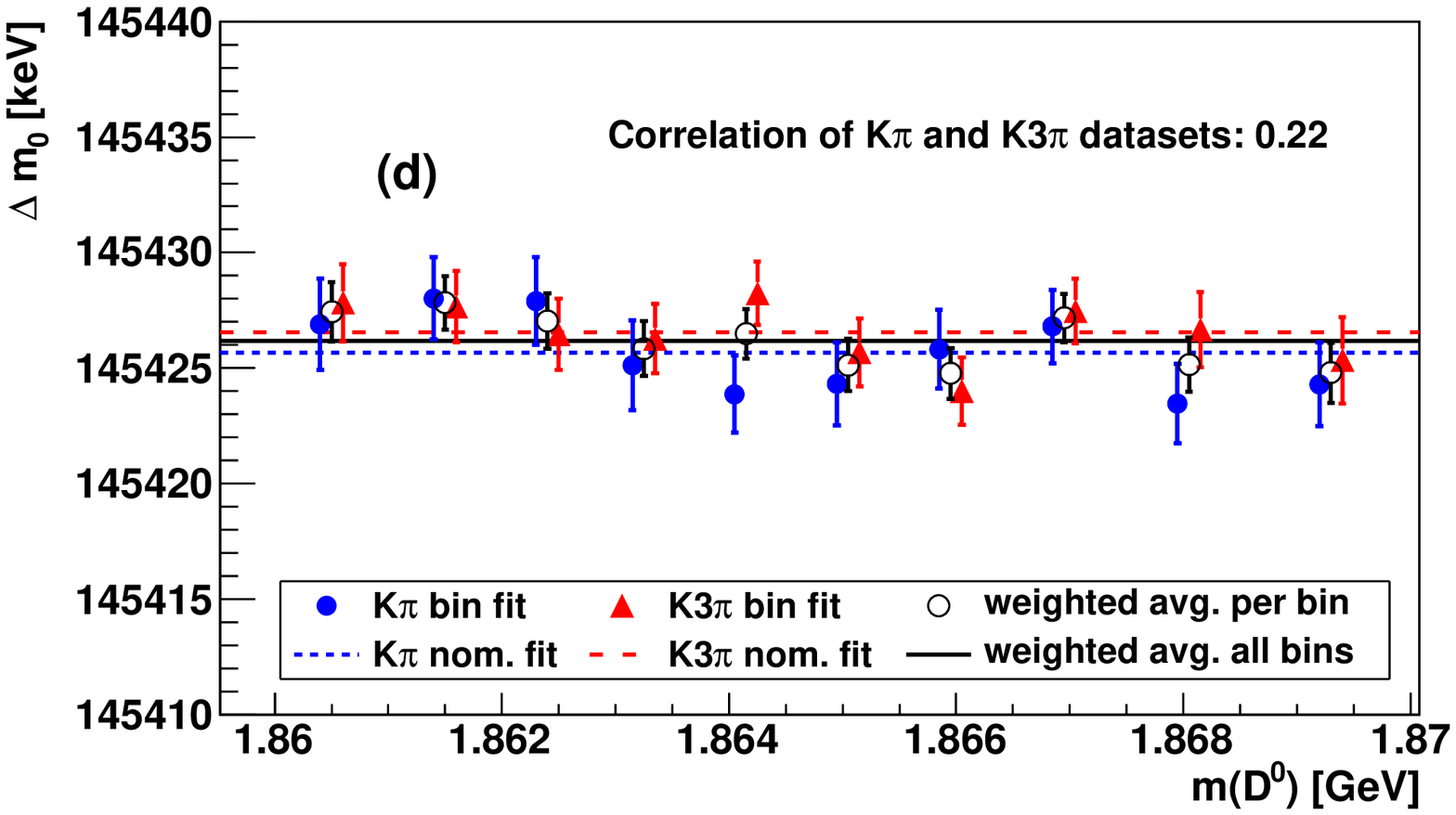} \label{fig:mslicerbw}} \\
   \subfigure{\includegraphics[scale=0.43]{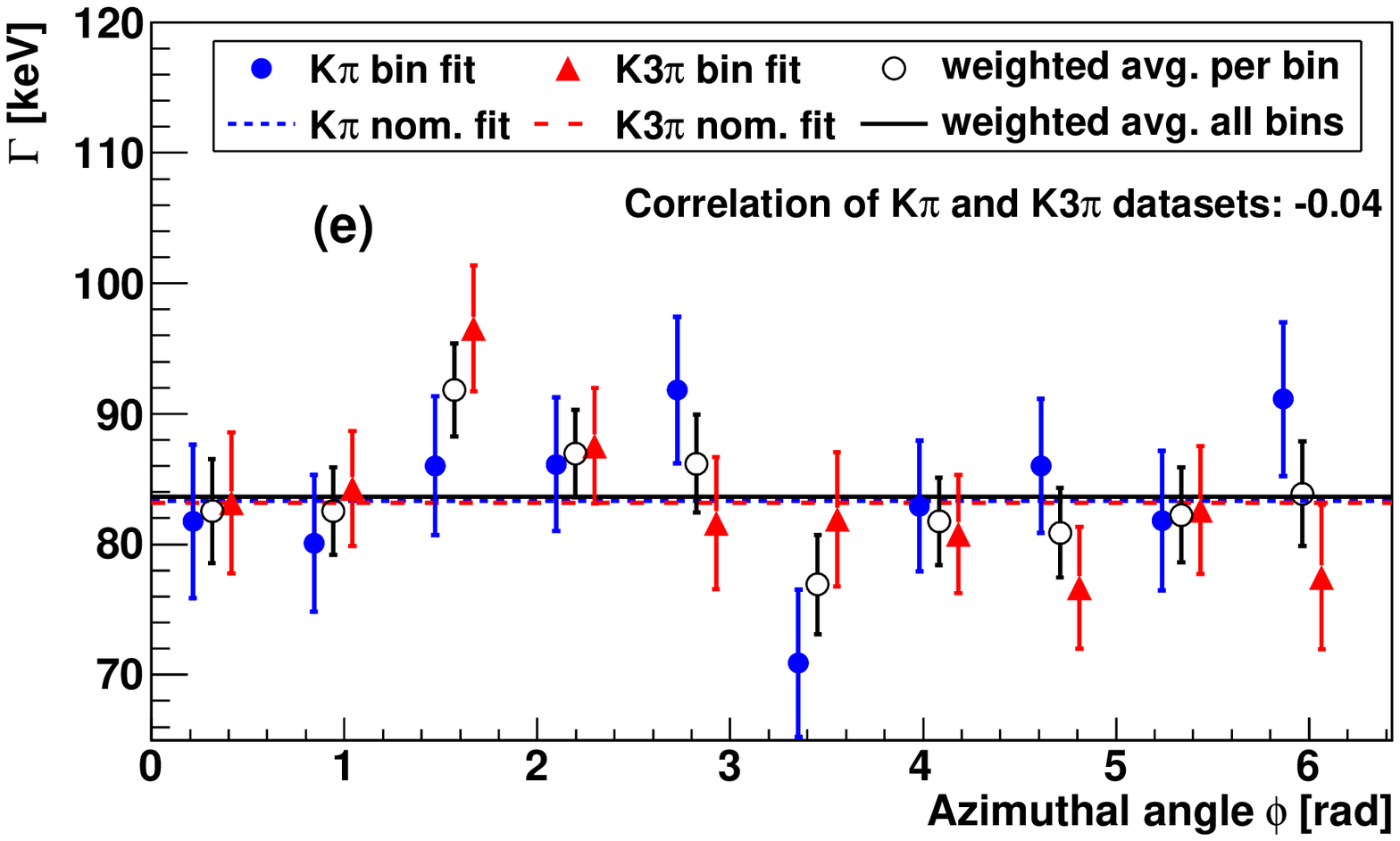} \label{fig:azwidth}} 
   \subfigure{\includegraphics[scale=0.43]{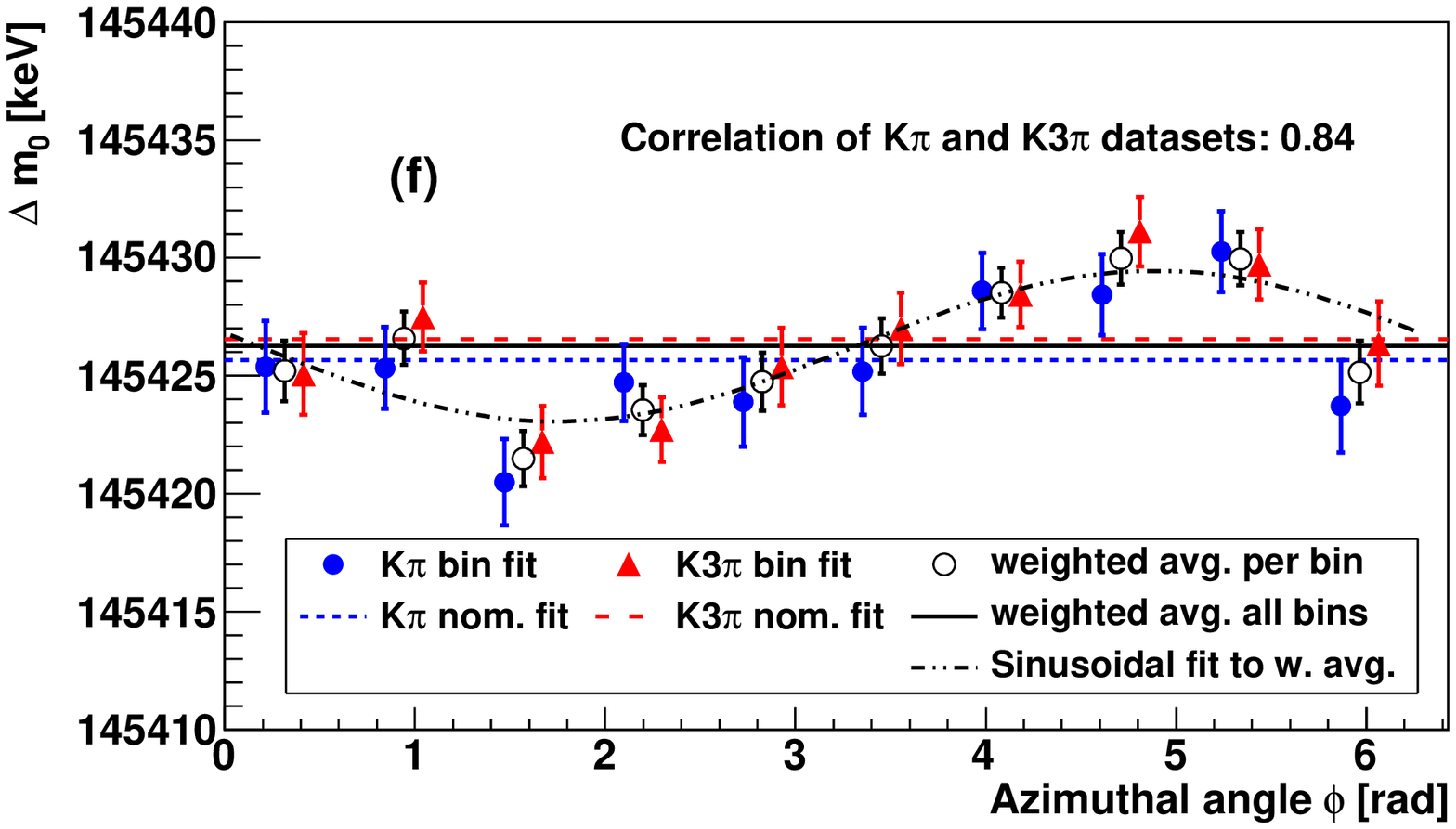} \label{fig:azrbw}} \\
   \end{center}
   \caption{(color online) The values of $\Gamma$ (left) and $\Delta m_0$ (right) obtained from fits to data divided into 10 disjoint subsets in laboratory momentum $p$ (top row), reconstructed $D^{0}$ mass (center row), and azimuthal angle (bottom row).  The quantities $p$ and $\phi$ are defined by the $D^{*+}$ momentum. Each point represents an individual fit and each horizontal line is the nominal fit result (i.e. integrating over the variable). The correlation value of $\Gamma$ (or $\Delta m_0$) measured from the $D^0\to K^-\pi^+$ and $D^0\to K^-\pi^+\pi^-\pi^+$ samples for each of the variables chosen is given above each plot. The widths from the nominal fits and the weighted average agree well and the corresponding lines are visually indistinguishable.}
   \label{fig:sys}
\end{figure*}

\subsection{\boldmath Systematics using disjoint subsets}
\label{sec:disjointsub}
We chose to carefully study laboratory momentum, reconstructed $D^0$ mass, and azimuthal angle $\phi$ in order to search for variations larger than those expected from statistical fluctuations. For each disjoint subset, we use the resolution function parameter values and $\Delta m_0$ offset determined from the corresponding MC subset.

If the fit results from the disjoint subsets are compatible with a constant value, in the sense that $\chi^2/\nu \leq 1$ where $\nu$ denotes the number of degrees of freedom, we assign no systematic uncertainty. However, if we find $\chi^2/\nu > 1$ and do not determine an underlying model which might be used to correct the data, we ascribe an uncertainty using a variation on the scale factor method used by the Particle Data Group (see the discussion of unconstrained averaging~\cite{ref:pdg2012}). The only sample which we do not fit to a constant is that for $\Delta m_0$ in intervals of azimuthal angle.  We discuss below how we estimate the associated systematic uncertainty.

In our version of this procedure, we determine a factor that scales the statistical uncertainty to the total uncertainty. The remaining uncertainty is ascribed to unknown detector issues and is used as a measure of systematic uncertainty according to

\begin{align}
\label{eq:sys}
\sigma_{\text{sys}}  &= \sigma_{\text{stat}} \sqrt{S^2 - 1} 
\end{align}
\noindent where the scale factor is defined as $S^2 = \chi^2/\nu$. The $\chi^2$ statistic gives a measure of fluctuations, including those expected from statistics, and those from systematic effects.  Once we remove the uncertainty expected from statistical fluctuations, we associate what remains with a possible systematic uncertainty. 

We expect that $ \chi^2 / \nu $ will have an average value of unity if there are no systematic uncertainties that distinguish one subset from another. If systematic deviations from one subset to another exist, then we expect that $\chi^2/\nu$ will be greater than unity. Even if there are no systematic variations from one disjoint subset to another, $ \chi^2 / \nu $ will randomly fluctuate above 1 about half of the time.  To be conservative, we assume that any observation of $ \chi^2 /  \nu > 1 $ originates from a systematic variation from one disjoint subset to another. This approach has two weaknesses. If used with a large number of subsets it could hide real systematic uncertainties.  For example, if instead of 10 subsets we chose 1000 subsets, the larger statistical uncertainties wash out any real systematic variation. Also, if used with a large number of variables, about half the disjoint sets will have upward statistical fluctuations, even in the absence of any systematic variation. We have chosen to use only three disjoint sets of events, and have divided each into 10 subsets to mitigate the effects of such problems.

We choose the range for each subset to have approximately equal statistical sensitivity. In each subset of each variable we repeat the full fit procedure (determine the resolution function from MC and fit data floating $\epsilon, \Gamma, \Delta m_0,$ and $c$). Figs.~\ref{fig:plabwidth} and~\ref{fig:plabrbw} show the fit results in subsets of laboratory momentum for $\Gamma$ and $\Delta m_0$, respectively. Neither $D^0$ mode displays a systematic pattern of variation; however, we assign small uncertainties for each channel using Eq.~(\ref{eq:sys}).  Similarly, Figs.~\ref{fig:mslicewidth} and~\ref{fig:mslicerbw} show the results in ranges of reconstructed $D^0$ mass for $\Gamma$ and $\Delta m_0$.  While neither mode displays an obvious systematic pattern of variation, the width for the $K^-\pi^+\pi^-\pi^+$ mode is assigned its largest uncertainty of $1.53 \kev$ using Eq.~(\ref{eq:sys}). 

Figures~\ref{fig:azwidth} and~\ref{fig:azrbw} show $\Gamma$ and $\Delta m_0$, respectively, in subsets of azimuthal angle. In this analysis we have observed sinusoidal variations in the mass values for $D^0 \rightarrow K^-\pi^+$, $D^0\rightarrow K^-\pi^+\pi^-\pi^+$, and $K_{S}^{0}\rightarrow \pi^+\pi^-$, so the clear sinusoidal variation of $\Delta m_0$ was anticipated. The important aspect for this analysis is that, for such deviations, the average value is unbiased by the variation in $\phi$. For example, the average value of the reconstructed $K_S^0$ mass separated into intervals of $\phi$ is consistent with the mass value integrating across the full range. The width plots do not display azimuthal dependencies, but each mode has $\chi^2/\nu > 1$ and is assigned a small systematic uncertainty using Eq.~(\ref{eq:sys}). The lack of sinusoidal variation of $\Gamma$ with respect to $\phi$ is notable because $\Delta m_0$ (which uses reconstructed $D$ masses) shows a clear sinusoidal variation. The results for the $D^0 \rightarrow K^-\pi^+$ and $D^0\rightarrow K^-\pi^+\pi^-\pi^+$ datasets are highly correlated, and shift together. The signs and phases of the variations of $\Delta m_0$ agree with those observed for $D^0 \rightarrow K^-\pi^+$, $D^0\rightarrow K^-\pi^+\pi^-\pi^+$, and $K_{S}^{0}\rightarrow \pi^+\pi^-$. We take half of the amplitude obtained from the sinusoidal fit shown on Fig.~\ref{fig:azrbw} as an estimate of the uncertainty. An extended investigation revealed that at least part of this dependence originates from small errors in the magnetic field from the map used in track reconstruction. There is some evidence that during the field mapping (see Ref.~\cite{ref:babar}) the propeller arm on which the probes were mounted flexed, which mixed the radial and angular components of the magnetic field.

\begin{figure*}[p]
   \begin{center}
   \subfigure{\includegraphics[scale=0.43]{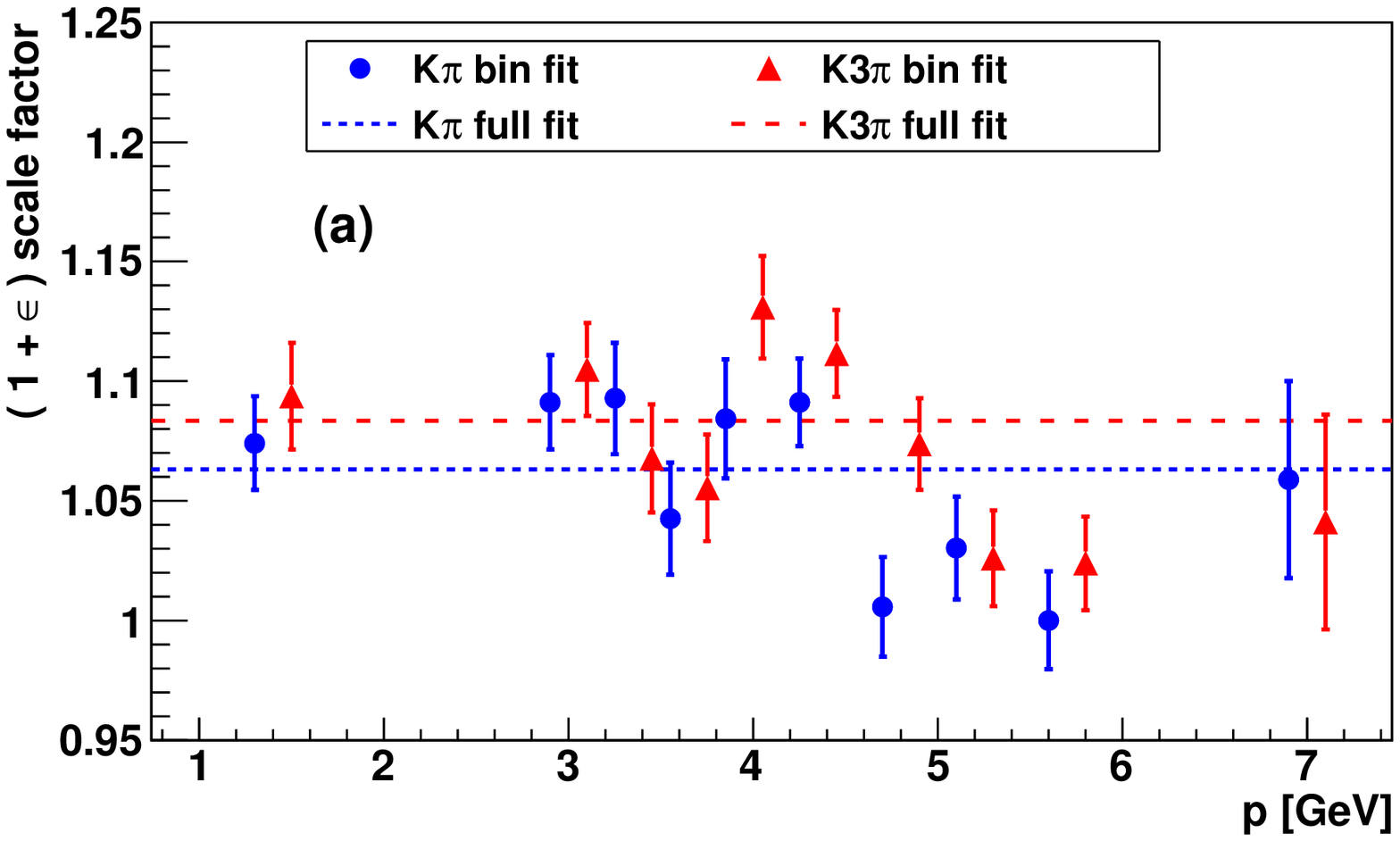} \label{fig:plabep}} 
   \subfigure{\includegraphics[scale=0.43]{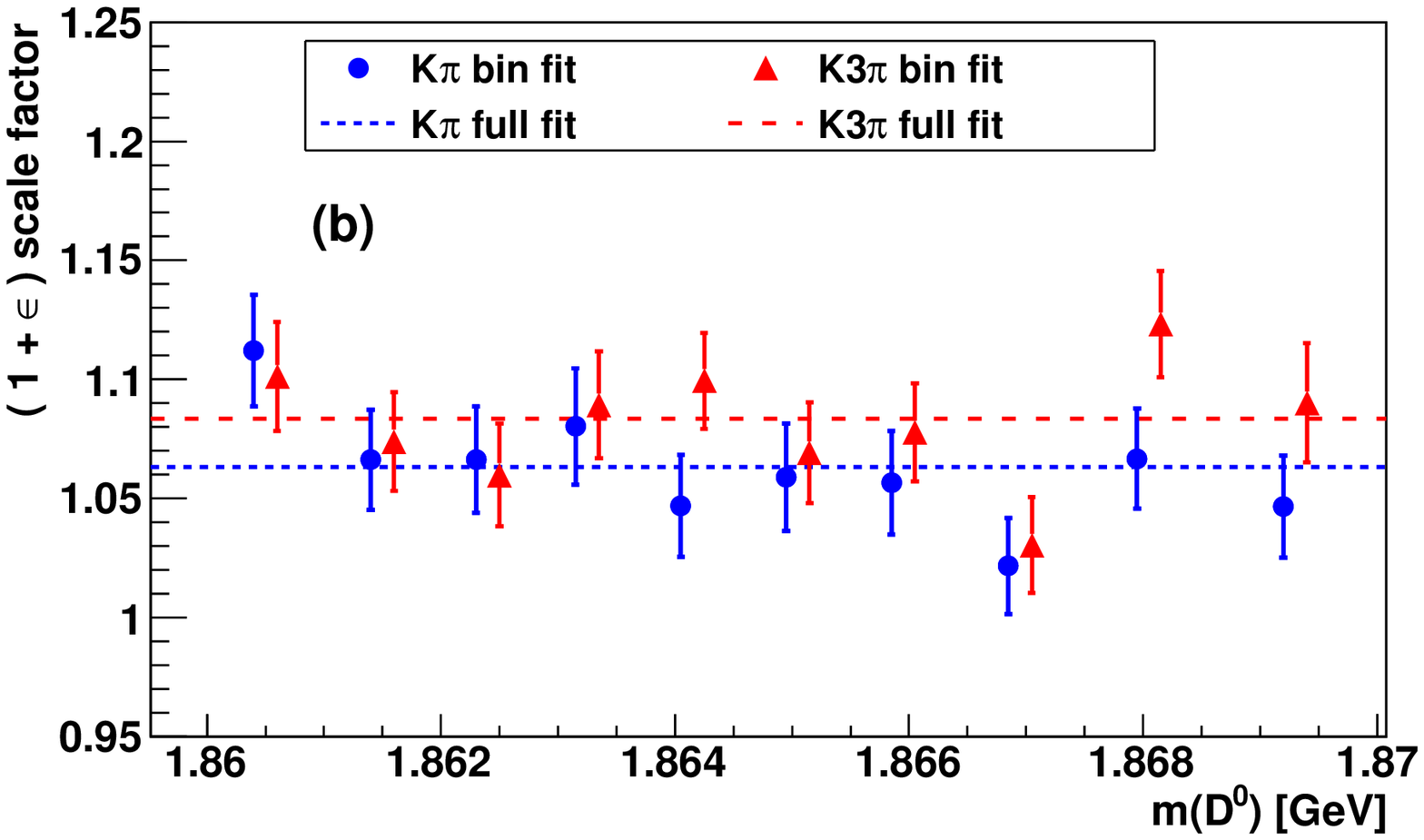} \label{fig:msliceep}} \\
   \subfigure{\includegraphics[scale=0.43]{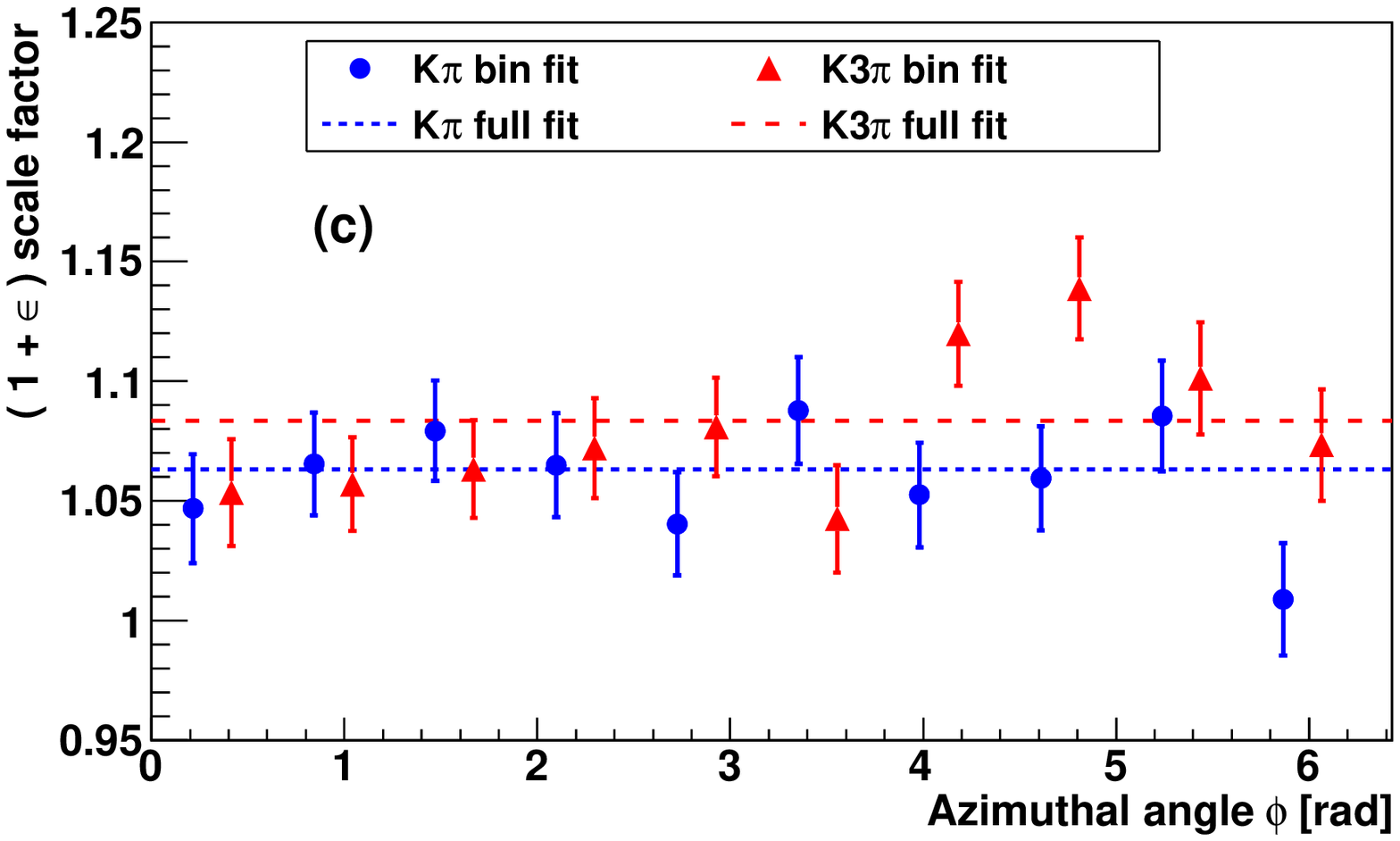} \label{fig:aziep}}
   \end{center}
   \caption{ (color online) The values of the scale factor $(1+\epsilon)$ obtained from fits to data divided into 10 disjoint subsets in laboratory momentum $p$, reconstructed $D^{0}$ mass, and azimuthal angle.  The quantities $p$ and $\phi$ are defined by the $D^{*+}$ laboratory momentum. Each point represents an individual fit and each horizontal line is the nominal fit result (i.e. integrating over the variable).}
   \label{fig:sysep}
\end{figure*}

The FWHM values of the resolution functions vary by about 8\% for each decay channel. For $D^0\rightarrow K^-\pi^+$ the FWHM ranges from $275 \kev$ to $325 \kev$ for the 30 disjoint subsets studied. The FWHM of the $D^0\rightarrow K^-\pi^+\pi^-\pi^+$ resolution function ranges are $310 \kev$ to $350 \kev$ for the 30 disjoint subsets studied. Fig.~\ref{fig:sysep} shows the values of the scale factor corresponding to the values of $\Gamma$ and $\Delta m_0$ shown in Fig.~\ref{fig:sys}.

\subsection{\boldmath Additional systematics}
We estimate the uncertainty associated with the correction parameters for the detector material model and magnetic field by examining the variation between the nominal parameter values and those obtained by tuning to the $m_{\text{PDG}}\left(K_{S}^{0}\right)\pm 1\sigma_{\text{PDG}}$ mass values~\cite{ref:pdg2012}. The width measured from the $D^0\to K^-\pi^+$ mode fluctuates equally around the value from the fit using the nominal correction parameters.  We take the larger of the differences and assign an uncertainty of $0.29 \kev$.  The value of $\Delta m_0$ for this mode fluctuates symmetrically around the nominal value and we assign an uncertainty of $0.75 \kev$. The width measured from the $D^0\to K^-\pi^+\pi^-\pi^+$ fluctuates asymmetrically around the nominal value, and we use the larger difference to assign an uncertainty of $0.18 \kev$. The value of $\Delta m_0$ for this mode fluctuates symmetrically around the nominal value, and we assign an uncertainty of $0.81 \kev$.

We use the Blatt-Weisskopf radius $r = 1.6 \gev^{-1}$ ($\sim0.3$ fm)~\cite{Schwartz:2002hh}. To estimate the systematic effect due to the choice of $r$ we refit the distributions treating the $D^{*+}$ as a point-like particle ($r=0$). We see a small shift of $\Gamma$, that we take as the estimate of the uncertainty, and an effect on the RBW pole position that is a factor of 100 smaller than the fit uncertainty, that we neglect.

We determine the systematic uncertainty associated with the resolution function by refitting the data with variations of its parametrization.  We take the covariance matrix from the fit to MC resolution samples for each mode (see Tables~\ref{tab:mcres_kpi_cov} and~\ref{tab:mcres_k3pi_cov} in the Appendix) and use it to generate 100 variations of these correlated Gaussian-distributed shape parameters.  We use these generated values to refit the data, and take the root-mean-squared (RMS) deviation of the resulting fit values as a measure of systematic uncertainty. This process implicitly accounts for the uncertainty associated with the reconstruction offset.

Our choice of fit range in $\Delta m$ is somewhat arbitrary, so we study the effect of systematically varying its end point by repeating the fit procedure every $1 \mev$ from the nominal fit end point, $\Delta m = 0.1665 \gev$, down to $\Delta m = 0.1605 \gev$. Altering the end point of the fit changes the events associated with the RBW tails and those associated with the continuum background. Each step down allows the background to form a different shape, which effectively estimates an uncertainty in the background parametrization. Values below $\Delta m = 0.16 \gev$ are too close to the signal region to provide a reasonable choice of end point. There is no clear way to estimate the associated systematic uncertainty, so we take the largest deviation from the nominal fit as a conservative estimate.

The shape of the background function in Eq.~(\ref{eq:bkgpdf}) is nominally determined only by the parameter $c$ and the residuals in Figs.~\ref{fig:rdfits_kpi} and~\ref{fig:rdfits_k3pi} show signs of curvature indicating possible systematic problems with the fits. Changing the end points over the range considered changes the values of $ c $ substantially from $-1.97$ to $-3.57$, and some fits remove all hints of curvature in the residuals plot. We also examine the influence of the background parametrization near threshold by changing $\sqrt{u}$ in Eq.~(\ref{eq:bkgpdf}) to $u^{0.45}$ and $u^{0.55}$. The value of the fractional power controls the shape of the background between the signal peak and threshold. For example, at $\Delta m = 0.142$ \gev changing the power from 0.5 to 0.45 and 0.55 varies the background function by +18\% and -15\%, respectively. The RBW pole position is unaffected by changing the background description near threshold while $\Gamma$ shifts symmetrically around its nominal values. We estimate the uncertainty due to the description of the background function near threshold by taking the average difference to the nominal result.

In the binned maximum likelihood fits we nominally choose an interval width of $50 \kev$.  As a systematic check, the interval width was halved and the fits to the data were repeated.  The measured $\Gamma$ and $\Delta m_0$ values for both modes are identical except for the width measured in the $D^0\to K^-\pi^+\pi^-\pi^+$ decay mode. We take the full difference as the systematic uncertainty for the choice of interval width.

\subsection{\boldmath Fit Validations}
\label{sec:validation}
We generate signal MC with $\Gamma = 88 \kev$ and $\Delta m_0 = 0.1454 \gev$.  The background is taken from a MC cocktail and paired with the signal in the same ratio as from the corresponding fits to data. Fits to both decay modes describe the validation samples well.  The fit results are summarized in Table~\ref{table:valsummary}.  We observe a small bias in the fitted width for the $D^0 \to K^-\pi^+\pi^-\pi^+$ mode. We take the full difference between the fitted and generated value of the width and assign a $1.5 \kev$ error. 

We also investigated the uncertainty due to radiative effects by examining the subset of these events generated without PHOTOS~\cite{Barberio1994291}. The values of the RBW pole are identical between the fits to the total validation signal MC sample and the subsets, so we do not assign a systematic uncertainty to the poles for radiative effects. The widths measured in each mode show a small difference to the results from the nominal validation sample. We take half of this difference as a conservative estimate of the systematic uncertainty associated with radiative effects.

\begin{table}[h]
\centering
\caption{Summary of results of the fits to the $D^0\to K^-\pi^+$ and $D^0\to K^-\pi^+\pi^-\pi^+$ validation MC samples. The width from the $D^0 \to K^-\pi^+\pi^-\pi^+$ decay mode has a small bias, which we take as a systematic uncertainty.}
\begin{tabular}{cc@{\hspace{2mm}}c@{\hspace{2mm}}c}
\hline \hline  \\[-1.7ex]
Fit value & Generated & $D^0\to K\pi$ & $D^0\to K\pi\pi\pi$ \\[-1.7ex] \\ \hline  \\[-1.7ex]
$\Gamma [\kev]$ & 88.0 & $88.5 \pm 0.8$ & $89.5 \pm 0.6$ \\ 
scale factor, $1+\epsilon$ & 1.0 & $1.003 \pm 0.004$ & $1.000 \pm 0.001$ \\ 
$\Delta m_0 [\kev]$ & 145400.0 & $145399.7 \pm 0.4$ & $145399.2 \pm 0.4$ \\ 
$\chi^2/\nu$ & -- & $613/540$ & $770/540$   \\[-1.7ex]\\ \hline\hline
\end{tabular}
\label{table:valsummary}
\end{table}

\subsection{\boldmath Determining correlations}
\label{sec:detcorr}
The fourth and seventh columns in Table~\ref{table:syswithcorr} list the correlations between the $D^0\to K^-\pi^+$ and $D^0\to K^-\pi^+\pi^-\pi^+$ systematic uncertainties.  These correlations are required to use information from both measurements to compute the average. The correlations in laboratory momentum, reconstructed $D^0$ mass, and azimuthal angle disjoint subsets are calculated by finding the correlation between the 10 subsets of $D^0\to K^-\pi^+$ and $D^0\to K^-\pi^+\pi^-\pi^+$ for each of the variables.  In a similar way we can construct datasets using the sets of correction parameters for magnetic field, detector material model, and the $\Delta m$ fit range. We assume no correlation for the resolution shape parameters and the validation shifts, which are based on the individual reconstructions. Our studies show that the values chosen for the Blatt-Weisskopf radius and interval width affect each mode identically, so we assume that they are completely correlated.

\subsection{\boldmath Consistency checks}
In addition to the investigations into the sources of systematic uncertainty, we also perform a number of consistency checks.  These checks are not used to assess systematics, nor are they included in the final measurements, but serve to reassure us that the experimental approach and fitting technique behave in reasonable ways.  First, we lower the $p^*$ cut from $3.6 \gev$ to $2.4 \gev$. This allows in more background and tracks with poorer resolution, but the statistics increase by a factor of three.  Correspondingly, the signal-to-background ratios measured at the peak and in the tails decrease by approximately a factor of three.  The fit results for this larger dataset are consistent with the nominal fit results.  The second consistency check widens the reconstructed $D^0$ mass window from $10 \mev$ to $30 \mev$.  Again, this increases the number of background events and improves statistical precision with central values that overlap with the nominal fit results.  Finally, we fix the scale factor in the fit to data to report statistical uncertainties on $\Gamma$ similar to those in the measurement by CLEO~\cite{PhysRevD.65.032003}.  Our reported ``statistical'' uncertainties on $\Gamma$ are from a fit in which $\epsilon$ floats.  As expected, there is a strong negative correlation between $\epsilon$ and $\Gamma$ with $\rho\left(\Gamma, \epsilon\right) \approx -0.85$. If less of the spread in the data is allotted to the resolution function then it must be allotted to the RBW width, $\Gamma$.  We refit the $D^0\to K^-\pi^+$ and $D^0\to K^-\pi^+\pi^-\pi^+$ samples fixing $\epsilon$ to the value from the fit where it was allowed to float.  This effectively maintains the same global minimum while decoupling the uncertainty in $\Gamma$ from $\epsilon$.  The statistical uncertainty on the width decreases from $1.7 \kev$ to $0.9 \kev$ for the $D^0\to K^-\pi^+$ decay mode and from $1.5 \kev$ to $0.8 \kev$ for the $D^0\to K^-\pi^+\pi^-\pi^+$ decay mode.



\section{\boldmath Combining results}
\label{sec:combmodes}
Using the correlations shown in Table~\ref{table:syswithcorr} and the formalism briefly outlined below, we determine the values for the combined measurement. For each quantity, $\Gamma$ and $\Delta m_0$, we have a measurement from the $D^0\to K^-\pi^+$ and $D^0\to K^-\pi^+\pi^-\pi^+$ modes. So, we start with a $2\times2$ covariance matrix

\begin{align}
\begin{split}
V &= \left( 
\begin{array}{cc}
\sigma_{K\pi}^2 & {\text{cov}}(K\pi, K\pi\pi\pi)\\
{\text{cov}}(K\pi, K\pi\pi\pi) & \sigma_{K\pi\pi\pi}^2 \end{array} \right) \\
&=\left(
\begin{array}{cc}
 \sigma_{K\pi, {\text{stat}}}^2 + \sigma_{K\pi, {\text{sys}}}^2 & \sum_{i}{\rho_i\, \sigma_{K\pi, i} \,\sigma_{K\pi\pi\pi, i}}\\
\sum_{i}{\rho_i \, \sigma_{K\pi, i}\, \sigma_{K\pi\pi\pi, i}} & \sigma_{K\pi\pi\pi, {\text{stat}}}^2 + \sigma_{K\pi\pi\pi, {\text{sys}}}^2 \end{array} \right)
\end{split}
\end{align}

\noindent where $i$ is an index which runs over the sources of systematic uncertainty. In the final step we expand the notation to explicitly show that the diagonal entries incorporate the full systematic uncertainty and that the statistical uncertainty for the individual measurements plays a part in determining the weights. The covariance matrices are calculated using Table~\ref{table:syswithcorr} and the individual measurements.  From the covariance matrix we extract the weights, $w$, for the best estimator of the mean and variance using $w_i = \sum_{k}{V^{-1}_{i k}}/\sum_{j k}{V^{-1}_{j k}}$:

\begin{equation}
w_{\Gamma} = \left( \begin{array}{cc}
w_{K\pi} \\
w_{K\pi\pi\pi}
\end{array}\right) 
= \left( \begin{array}{cc}
0.650 \\
0.350
\end{array}\right)
\end{equation}

\begin{equation}
w_{\Delta m_0} = \left( \begin{array}{cc}
w_{K\pi} \\
w_{K\pi\pi\pi}
\end{array}\right) 
= \left( \begin{array}{cc}
0.672\\
0.328\end{array}\right).
\end{equation}

\noindent The weights show that the combined measurement is dominated by the cleaner $D^0\to K^-\pi^+$ mode. The total uncertainty can be expressed as

\begin{align}
\begin{split}
\sigma^2 = &\sum_{i=1,2}{\left(w_{i} \sigma_{\text{stat}, i}\right)^2} \\ 
&+ \sum_{i=1,2}{\left(w_{i} \sigma_{\text{sys}, i}\right)^2} + 2 w_1 w_2 \sum_{j=1,11}{\rho_{j} \sigma^{K\pi}_{\text{sys}, j}  \sigma^{K\pi\pi\pi}_{\text{sys}, j}}.
\end{split}
\label{eq:combo_statsys}
\end{align}

\noindent The statistical contribution is the first term and is simply calculated using the individual measurements and the weights.  The remaining two terms represent the systematic uncertainty, which is simply the remainder of the total uncertainty after the statistical contribution has been subtracted. The weighted results are $\Gamma = \left(83.3 \pm 1.2 \pm 1.4\right) \kev$ and $\Delta m_0 = \left(145\,425.9 \pm 0.4 \pm 1.7\right) \kev$.

\section{Summary and conclusions}
\label{sec:conclusion}

We have measured the pole mass and the width of the $D^{*+}$ meson with unprecedented precision, analyzing a high-purity sample of continuum-produced $D^{*+}$ in \epem\ collisions at approximately $10.6 \gev$, equivalent to approximately $477 \invfb$, collected by the \babar\ detector. The results for the two independent $ D^0 $ decay modes agree with each other well.  The dominant systematic uncertainty on the RBW pole position comes from the azimuthal variation.  For the decay mode $D^0\rightarrow K^-\pi^+$ we obtain $\Gamma = \left(83.4 \pm 1.7 \pm 1.5\right) \kev$ and $\Delta m_0 =  \left(145\,425.6 \pm 0.6 \pm 1.7\right) \kev$ while for the decay mode $D^0\rightarrow K^-\pi^+\pi^-\pi^+$ we obtain $\Gamma = \left(83.2 \pm 1.5 \pm 2.6\right) \kev$ and $\Delta m_0 = \left(145\,426.6 \pm 0.5 \pm 1.9\right) \kev$. Accounting for correlations, we obtain the combined measurement values $\Gamma = \left(83.3 \pm 1.2 \pm 1.4\right) \kev$ and $\Delta m_0 = \left(145\,425.9 \pm 0.4 \pm 1.7\right) \kev$.

The experimental value of $g_{D^*D\pi}$ is calculated using the relationship between the width and the coupling constant,

\begin{align}
\Gamma &= \Gamma\left(D^{0}\pi^+\right) + \Gamma\left(D^{+}\pi^0\right) + \Gamma\left(D^{+}\gamma\right) \\
&\approx \Gamma\left(D^{0}\pi^+\right) + \Gamma\left(D^{+}\pi^0\right) \\
 &\approx \frac{g^2_{D^* D^0\pi^+}}{24\pi m^2_{D^{*+}}} p^3_{\pi^+} + \frac{g^2_{D^* D^+\pi^0}}{24\pi m^2_{D^{*+}}} p^3_{\pi^0}
\end{align}

\noindent where we have again ignored the electromagnetic contribution. The strong couplings can be related through isospin by $g_{D^*D^0\pi^+} = -\sqrt{2} g_{D^* D^+\pi^0}$~\cite{PhysRevD.65.032003}. Using $\Gamma$ and the mass values from Ref.~\cite{ref:pdg2012} we determine the experimental coupling $g_{D^*D^0\pi^+}^{\text{exp}} = 16.92 \pm 0.13 \pm 0.14$. The universal coupling is directly related to the strong coupling by $\hat{g} = g_{D^*D^0\pi^+} f_\pi / \left(2 \sqrt{m_{D}m_{D^*}}\right)$. This parametrization is different from that of Ref.~\cite{PhysRevD.65.032003} and is chosen to match a common choice when using chiral perturbation theory, as in Refs.~\cite{PhysRevC.83.025205, PhysRevD.66.074504}. With this relation and $f_\pi = 130.41 \mev$, we find $\hat{g}^{\text{exp}} = 0.570 \pm 0.004 \pm 0.005$. 

\begin{table}
\begin{center}
\caption{Selected rows from Table 11 of Ref.~\cite{PhysRevD.64.114004}.  State names correspond to the current PDG listings. The third column is the ratio, $R= \Gamma/\hat{g}^2$, extracted from the model in Ref.~\cite{PhysRevD.64.114004}. The values of $\hat{g}$ were obtained from the data available in 2001.}
\begin{tabular}{c@{\hspace{3mm}}ccc}
\hline\hline  \\[-1.7ex]
\multirow{2}{*}{State} & \multirow{2}{*}{Width ($\Gamma$)} & $R$ & \multirow{2}{*}{$\hat{g}$} \\
& & (model) &  \\[-1.7ex]\\ \hline  \\[-1.7ex]
$D^{*}\left(2010\right)^{+}$ & $96 \pm 4 \pm 22 \kev$ & $143 \kev$ & $0.82 \pm 0.09$ \\ 
$D_{1}\left(2420\right)^{0}$ & $18.9^{+4.6}_{-3.5} \mev$ & $16 \mev$ & $1.09 ^{+0.12}_{-0.11}$ \\ 
$D_{2}^{*}\left(2460\right)^{0}$ &$23 \pm 5 \mev$ & $38 \mev$ & $0.77 \pm 0.08$  \\[-1.7ex]\\ \hline \hline
\end{tabular}
\label{table:eichten}
\end{center}
\end{table}

The paper by Di Pierro and Eichten~\cite{PhysRevD.64.114004} quotes results in terms of a ratio, $R = \Gamma/\hat{g}^2$, which involves the width of the particular state and provides a straightforward method for calculating the corresponding value of the universal coupling constant within their model. The coupling constant should then take the same value for the selected $D^{\left(*\right)}$ decay channels listed in Table~\ref{table:eichten}, which shows the values of the ratio $R$ extracted from the model and the experimental values for $\Gamma$, as they were in 2001. At the time of publication, $\hat{g}$ was consistent for all of the modes in Ref.~\cite{PhysRevD.64.114004}. In 2010, \babar\ published much more precise results for the $D_1\left(2420\right)^0$ and $D_2^*\left(2460\right)^0$~\cite{PhysRevD.82.111101}. Using those results, this measurement of $\Gamma$, and the ratios from Table~\ref{table:eichten}, we calculate new values for the coupling constant $\hat{g}$.  Table~\ref{table:zach_eichten} shows the updated results.  We estimate the uncertainty on the coupling constant value assuming $\sigma_{\Gamma} \ll \Gamma$.  The updated widths reveal significant differences among the extracted values of $\hat{g}$.

\begin{table}
\begin{center}
\caption{Updated coupling constant values using the latest width measurements. Ratio values are taken from Table~\ref{table:eichten}.  Significant differences are seen among the coupling constants calculated using the updated width measurements.}
\begin{tabular}{c@{\hspace{3mm}}ccc}
\hline \hline \\[-1.7ex] 
\multirow{2}{*}{State} & \multirow{2}{*}{Width ($\Gamma$)} & $R$ & \multirow{2}{*}{$\hat{g}$} \\
& & (model) & \\ \hline  \\[-1.7ex] 
$D^{*}\left(2010\right)^{+}$ & $83.3 \pm 1.2 \pm 1.4 \kev$ & $143 \kev$ & $0.76 \pm 0.01$ \\ 
$D_{1}\left(2420\right)^{0}$ & $31.4 \pm 0.5 \pm 1.3 \mev$ & $16 \mev$ & $1.40 \pm 0.03$ \\ 
$D_{2}^{*}\left(2460\right)^{0}$ &$50.5 \pm 0.6 \pm 0.7 \mev$ & $38 \mev$ & $1.15 \pm 0.01$ \\[-1.7ex]  \\ \hline \hline
\end{tabular}
\label{table:zach_eichten}
\end{center}
\end{table}

After completing this analysis, we became aware of Rosner's 1985 prediction that the $D^{*+}$ natural line width should be $83.9 \kev$ \cite{Rosner:1985dx}.  He calculated this assuming a  single quark transition model to use P-wave $K^* \to K\pi$ decays to predict P-wave $D^* \to D\pi$ decay properties. Although he did not report an error estimate for this calculation in that work, his central value falls well within our experimental precision. Using the same procedure and current measurements, the prediction becomes $(80.5 \pm 0.1) \kev$~\cite{rosnerPrivate}. A new lattice gauge calculation yielding  $\Gamma( D^{*+} ) = ( 76 \pm 7{}^{+8}{}_{-10} ) $ keV, has also been reported recently~\cite{Becirevic201394}.

The order of magnitude increase in precision confirms the observed inconsistency between the measured $D^{*+}$ width and the chiral quark model calculation by Di Pierro and Eichten~\cite{PhysRevD.64.114004}.  The precise measurements of the widths presented in Table~\ref{table:zach_eichten} provide solid anchor points for future calculations.


\section{Acknowledgments}
We are grateful for the 
extraordinary contributions of our \pep2\ colleagues in
achieving the excellent luminosity and machine conditions
that have made this work possible.
The success of this project also relies critically on the 
expertise and dedication of the computing organizations that 
support \babar.
The collaborating institutions wish to thank 
SLAC for its support and the kind hospitality extended to them. 
This work is supported by the
US Department of Energy
and National Science Foundation, the
Natural Sciences and Engineering Research Council (Canada),
the Commissariat \`a l'Energie Atomique and
Institut National de Physique Nucl\'eaire et de Physique des Particules
(France), the
Bundesministerium f\"ur Bildung und Forschung and
Deutsche Forschungsgemeinschaft
(Germany), the
Istituto Nazionale di Fisica Nucleare (Italy),
the Foundation for Fundamental Research on Matter (The Netherlands),
the Research Council of Norway, the
Ministry of Education and Science of the Russian Federation, 
Ministerio de Ciencia e Innovaci\'on (Spain), and the
Science and Technology Facilities Council (United Kingdom).
Individuals have received support from 
the Marie-Curie IEF program (European Union) and the A. P. Sloan Foundation (USA). 
The University of Cincinnati is gratefully acknowledged for its support of 
this research through a WISE (Women in Science and Engineering) fellowship to C. Fabby.

\bibliography{refs_bad2523.bib}
\appendix*
\section{}
\label{app:fitresults}

In this appendix we present the covariance and correlation matrices for the fits described in Sect.~\ref{sec:resfit} and~\ref{sec:datafit}.

\begin{table*}
\caption{Summary of the results from the fits to the MC resolution sample for the $D^0\to K^-\pi^+$ and $D^0\to K^-\pi^+\pi^-\pi^+$ channels (statistical uncertainties only). Parameters are defined in Eqs.~(\ref{eq:respdf}) and~(\ref{eq:resng}).}
\begin{tabular}{ccc}
\hline\hline \\[-1.7ex]
Parameter & $D^0\to K^-\pi^+$ & $D^0\to K^-\pi^+\pi^-\pi^+$ \\[-1.7ex] \\ \hline \\[-1.7ex]
 $f_{NG}$&   $0.00559 \pm 0.00018$ & $0.0054 \pm 0.00016$ \\
$\alpha$ &    $1.327 \pm 0.091$ & $1.830 \pm 0.092$ \\
$q$  &  $-23.04 \pm 1.02$ & $-29.24 \pm 1.07$ \\
$f_1$ & $0.640 \pm 0.013$ & $0.730 \pm 0.008$ \\
$f_2$ & $0.01874 \pm 0.00086$ & $0.02090 \pm 0.00069$ \\
$\mu_1\, (\kev)$ & $145402.36 \pm 0.33$ & $145402.84 \pm 0.24$ \\
$\mu_2\, (\kev)$ &  $145465.37 \pm 9.39$ & $145451.63 \pm 7.83$\\
$\mu_3\, (\kev)$ &  $145404.58 \pm 0.75$ & $145399.07 \pm 0.81$ \\
$\sigma_1\, (\kev)$& $119.84 \pm 0.84$ & $112.73 \pm 0.52$\\
$\sigma_2\, (\kev)$ & $722.89 \pm 20.6$ & $695.04 \pm 15.75$\\
$\sigma_3\, (\kev)$& $212.31 \pm 2.42$ & $209.54 \pm 2.41$  \\ \\[-1.7ex] \hline \hline
\end{tabular}
\label{tab:mcres}
\end{table*}

\begin{table*}
\caption{Covariance matrix for the parameters from the fit to $D^0 \to K^-\pi^+ $ MC resolution sample. Parameters are defined in Eqs.~(\ref{eq:respdf}) and~(\ref{eq:resng}). Symmetric elements are suppressed.}
\begin{tabular}{cccccccccccc}
\hline\hline \\[-1.7ex]
 & $f_{NG}$ & $\alpha$ & $q$ & $f_1$ & $f_2$ & $\mu_1$ & $\mu_2$ & $\mu_3$ & $\sigma_1$ &$\sigma_2$ &$\sigma_3$ \\[-1.7ex] \\ \hline \\[-1.7ex]
 $f_{NG}$&    \phantom{-}3.263e-08  &&&&&&&&&& \\
$\alpha$ &    \phantom{-}1.002e-05 & \phantom{-}8.311e-03  &&&&&&&&& \\
$q$  &  -1.139e-04 &-8.914e-02&  \phantom{-}1.033e+00  &&&&&&&& \\
$f_1$ & -7.780e-07& -3.250e-04&  \phantom{-}3.662e-03 & \phantom{-}1.581e-04  &&&&&&& \\
$f_2$ &  \phantom{-}5.671e-08 & \phantom{-}2.336e-05 &-2.627e-04 &-6.724e-06 & \phantom{-}5.761e-07  &&&&&& \\
$\mu_1$&    \phantom{-}1.064e-13 &-2.634e-11 &-4.741e-10 & \phantom{-}1.426e-10 & -3.353e-12 & \phantom{-}1.081e-13 &&&&& \\ 
$\mu_2$ &   -1.998e-10 &-1.059e-07 & \phantom{-}9.350e-07&  \phantom{-}2.265e-08& -1.913e-09 & \phantom{-}2.996e-13 & \phantom{-}8.823e-11 &&&& \\
$\mu_3$ &   -1.016e-11& -3.919e-09 & \phantom{-}4.775e-08 & \phantom{-}1.158e-09 &-6.553e-11 &-1.423e-13 &-1.102e-12 & \phantom{-}5.624e-13 &&& \\
$\sigma_1$ &   -4.662e-11& -1.949e-08  &\phantom{-}2.196e-07 & \phantom{-}1.012e-08 &-3.980e-10&  \phantom{-}9.854e-15 & \phantom{-}1.342e-12 & \phantom{-}7.143e-14  &\phantom{-}7.072e-13 && \\
$\sigma_2$ &  -2.474e-09& -1.035e-06&  \phantom{-}1.173e-05&  \phantom{-}1.584e-07 &-1.306e-08&  \phantom{-}1.144e-14 & \phantom{-}4.486e-11 & \phantom{-}1.887e-12 & \phantom{-}9.422e-12 &  \phantom{-}4.260e-10 & \\
$\sigma_3$ &  -1.756e-10 & -7.341e-08 & 8.275e-07 & \phantom{-}2.942e-08 &-1.469e-09 & \phantom{-}2.487e-14 & \phantom{-}5.008e-12 & \phantom{-}2.302e-13 & \phantom{-}1.818e-12  &\phantom{-}3.528e-11 &  \phantom{-}5.872e-12   \\ \\[-1.7ex] \hline \hline
\end{tabular}
\label{tab:mcres_kpi_cov}
\end{table*}

\begin{table*}
\caption{Parameter correlation coefficients for  the parameters from the fit to $D^0 \to K^-\pi^+ $ MC resolution sample. Parameters are defined in Eqs.~(\ref{eq:respdf}) and~(\ref{eq:resng}). Symmetric elements are suppressed.}
\begin{tabular}{cccccccccccc}
\hline\hline \\[-1.7ex]
 & $f_{NG}$ & $\alpha$ & $q$ & $f_1$ & $f_2$ & $\mu_1$ & $\mu_2$ & $\mu_3$ & $\sigma_1$ &$\sigma_2$ &$\sigma_3$ \\[-1.7ex] \\ \hline \\[-1.7ex]
 $f_{NG}$&    \phantom{-}1.000  &&&&&&&&&& \\
$\alpha$ &       \phantom{-}0.608 & \phantom{-}1.000 &&&&&&&&& \\
$q$  &      -0.621& -0.962&  \phantom{-}1.000&&&&&&&& \\
$f_1$ &    -0.343& -0.284&  \phantom{-}0.287&  \phantom{-}1.000 &&&&&&&\\
$f_2$ &   \phantom{-}0.414&  \phantom{-}0.338& -0.340& -0.705&  \phantom{-}1.000 &&&&&&\\
$\mu_1$&     \phantom{-}0.002& -0.001& -0.001&  \phantom{-}0.034& -0.013&  \phantom{-}1.000&&&&&\\
$\mu_2$ &    -0.118& -0.124&  \phantom{-}0.098&  \phantom{-}0.192& -0.268&  \phantom{-}0.097&  \phantom{-}1.000&&&& \\
$\mu_3$ &    -0.075& -0.057&  \phantom{-}0.063&  \phantom{-}0.123& -0.115& -0.577& -0.156&  \phantom{-}1.000&&& \\
$\sigma_1$ &   -0.307& -0.254&  \phantom{-}0.257&  \phantom{-}0.958& -0.624&  \phantom{-}0.036&  \phantom{-}0.170&  \phantom{-}0.113&  \phantom{-}1.000 && \\
$\sigma_2$ &   -0.664& -0.550&  \phantom{-}0.559&  \phantom{-}0.611& -0.834&  \phantom{-}0.002&  \phantom{-}0.231&  \phantom{-}0.122&  \phantom{-}0.543&  \phantom{-}1.000 & \\
$\sigma_3$ &   -0.401& -0.332&  \phantom{-}0.336&  \phantom{-}0.966& -0.799&  \phantom{-}0.031&  \phantom{-}0.220&  \phantom{-}0.127&  \phantom{-}0.892&  \phantom{-}0.705&  \phantom{-}1.000 \\ \\[-1.7ex] \hline \hline
\end{tabular}
\label{tab:mcres_kpi_corr}
\end{table*}

\begin{table*}
\caption{ Covariance matrix for the parameters from the fit to $D^0 \to K^-\pi^+ \pi^-\pi^+$ MC resolution sample. Parameters are defined in Eqs.~(\ref{eq:respdf}) and~(\ref{eq:resng}). Symmetric elements are suppressed.}
\begin{tabular}{cccccccccccc}
\hline\hline \\[-1.7ex]
 & $f_{NG}$ & $\alpha$ & $q$ & $f_1$ & $f_2$ & $\mu_1$ & $\mu_2$ & $\mu_3$ & $\sigma_1$ &$\sigma_2$ &$\sigma_3$ \\[-1.7ex] \\ \hline \\[-1.7ex]
 $f_{NG}$&    \phantom{-}2.746e-08 &&&&&&&&&& \\
$\alpha$ &    \phantom{-}9.170e-06  &\phantom{-}8.565e-03 &&&&&&&&& \\
$q$  &  -1.076e-04 &-9.539e-02  &\phantom{-}1.149e+00 &&&&&&&& \\
$f_1$ &  -3.981e-07& -1.799e-04&  \phantom{-}2.071e-03&  \phantom{-}6.953e-05 &&&&&&& \\
$f_2$ &  \phantom{-}4.133e-08&  \phantom{-}1.829e-05& -2.100e-04& -3.847e-06&  \phantom{-}4.784e-07 &&&&&& \\
$\mu_1$&    \phantom{-}1.274e-12&  \phantom{-}5.343e-10& -6.776e-09& -1.097e-10&  \phantom{-}9.246e-12&  \phantom{-}5.648e-14 &&&&& \\ 
$\mu_2$ &   -1.434e-10& -7.936e-08&  6.757e-07&  1.332e-08& -1.478e-09&  \phantom{-}1.399e-13& \phantom{-}6.134e-11 &&&& \\
$\mu_3$ &   -1.909e-13&  \phantom{-}2.382e-10&  2.094e-09& -6.916e-10&  \phantom{-}1.981e-11& -1.016e-13& -1.394e-12&  \phantom{-}6.582e-13 &&& \\
$\sigma_1$ &   -2.191e-11& -9.918e-09&  \phantom{-}1.142e-07&  \phantom{-}4.099e-09& -2.061e-10& -5.895e-15&  \phantom{-}7.264e-13& -4.344e-14&  \phantom{-}2.724e-13 && \\
$\sigma_2$ &  -1.669e-09& -7.535e-07&  \phantom{-}8.781e-06&  \phantom{-}7.332e-08& -8.820e-09& -2.122e-13&  \phantom{-}2.902e-11& -1.152e-13&  \phantom{-}3.967e-12&  \phantom{-}2.480e-10 & \\
$\sigma_3$ &  -1.428e-10& -6.452e-08&  \phantom{-}7.441e-07&  \phantom{-}1.919e-08& -1.303e-09& -3.679e-14&  \phantom{-}4.432e-12& -1.616e-13&  \phantom{-}1.084e-12&  \phantom{-}2.561e-11&  5.806e-12  \\ \\[-1.7ex] \hline \hline
\end{tabular}
\label{tab:mcres_k3pi_cov}
\end{table*}

\begin{table*}
\caption{Parameter correlation coefficients for the parameters from the fit to $D^0 \to K^-\pi^+ \pi^-\pi^+$ MC resolution sample. Parameters are defined in Eqs.~(\ref{eq:respdf}) and~(\ref{eq:resng}). Symmetric elements are suppressed.}
\begin{tabular}{cccccccccccc}
\hline\hline \\[-1.7ex]
 & $f_{NG}$ & $\alpha$ & $q$ & $f_1$ & $f_2$ & $\mu_1$ & $\mu_2$ & $\mu_3$ & $\sigma_1$ &$\sigma_2$ &$\sigma_3$ \\[-1.7ex] \\ \hline \\[-1.7ex]
 $f_{NG}$&    \phantom{-}1.000  &&&&&&&&&& \\
$\alpha$ &         \phantom{-}0.598 & \phantom{-}1.000 &&&&&&&&& \\
$q$  &      -0.606 &-0.962 &  \phantom{-}1.000&&&&&&&& \\
$f_1$ &        -0.288& -0.233 &  \phantom{-}0.232&  \phantom{-}1.000 &&&&&&&\\
$f_2$ &  \phantom{-}0.361&  \phantom{-}0.286& -0.283 & -0.667 &\phantom{-}1.000 &&&&&&\\
$\mu_1$&        \phantom{-}0.032&  \phantom{-}0.024& -0.027& -0.055&  \phantom{-}0.056&  \phantom{-}1.000&&&&&\\
$\mu_2$ &         -0.110& -0.109&  \phantom{-}0.080&  \phantom{-}0.204& -0.273&  \phantom{-}0.075&  \phantom{-}1.000&&&& \\
$\mu_3$ &        -0.001& \phantom{-}0.003& \phantom{-}0.002& -0.102&  \phantom{-}0.035& -0.527& -0.219&  \phantom{-}1.000 &&& \\
$\sigma_1$ &        -0.253& -0.205&  \phantom{-}0.204&  \phantom{-}0.942& -0.571& -0.048&  \phantom{-}0.178& -0.103&  \phantom{-}1.000 && \\
$\sigma_2$ &       -0.639& -0.517&  \phantom{-}0.520&  \phantom{-}0.558& -0.810& -0.057&  \phantom{-}0.235& -0.009&  \phantom{-}0.483&  \phantom{-}1.000 & \\
$\sigma_3$ &       -0.358& -0.289&  \phantom{-}0.288&  \phantom{-}0.955& -0.782& -0.064&  \phantom{-}0.235& -0.083&  \phantom{-}0.862&  \phantom{-}0.675&  \phantom{-}1.000 \\ \\[-1.7ex] \hline \hline
\end{tabular}
\label{tab:mcres_k3pi_corr}
\end{table*}

\begin{table*}
\caption{Covariance matrix for the parameters from the fit to $D^0 \to K^-\pi^+$ data. Parameters are defined in Eqs.~(\ref{eq:sigpdf}) and ~(\ref{eq:bkgpdf}). Symmetric elements are suppressed.}
\begin{tabular}{ccccccc}
\hline\hline \\[-1.7ex]
 & $\Delta m_0$ & $\epsilon$ & $N_{sig}$ & $N_{bkg}$ & $c$ & $\Gamma$\\[-1.7ex] \\ \hline \\[-1.7ex]
$\Delta m_0$ &  \phantom{-}3.181e-13 & &&&& \\
$\epsilon$ &  \phantom{-}4.060e-10  & \phantom{-}4.909e-05 & &&& \\
$N_{sig}$ &  \phantom{-}3.782e-06 & \phantom{-}3.533e-01  &\phantom{-}1.199e+04 &&&\\
$N_{bkg}$ & -3.692e-06 &-3.448e-01 &-8.631e+03 & \phantom{-}1.470e+05 & &\\
$c$ & -6.288e-09 &-5.534e-04 &-1.711e+01 & \phantom{-}1.668e+01 & \phantom{-}7.936e-02  &\\
$\Gamma$ & -1.017e-13 &-9.965e-09 &-1.084e-04 & \phantom{-}1.058e-04 & \phantom{-}1.779e-07 & \phantom{-}2.920e-12 \\ \\[-1.7ex] \hline \hline
\end{tabular}
\label{tab:rd_kpi_cov}
\end{table*}

\begin{table*}
\caption{Parameter correlation coefficients for the parameters from the fit to $D^0 \to K^-\pi^+$ data. Parameters are defined in Eqs.~(\ref{eq:sigpdf}) and ~(\ref{eq:bkgpdf}). Symmetric elements are suppressed.}
\begin{tabular}{ccccccc}
\hline\hline \\[-1.7ex]
 & $\Delta m_0$ & $\epsilon$ & $N_{sig}$ & $N_{bkg}$ & $c$ & $\Gamma$ \\[-1.7ex] \\ \hline \\[-1.7ex]
$\Delta m_0$ &  \phantom{-}1.000 & &&&&\\
$\epsilon$ &  \phantom{-}0.103  &\phantom{-}1.000  &&&&\\
$N_{sig}$ &  \phantom{-}0.061&  \phantom{-}0.461&  \phantom{-}1.000& &&\\
$N_{bkg}$ & -0.017 &-0.128& -0.206&  \phantom{-}1.000&  & \\
$c$ & -0.040& -0.280& -0.555&  \phantom{-}0.154&  \phantom{-}1.000&  \\
$\Gamma$ & -0.106& -0.832& -0.579&  \phantom{-}0.161&  \phantom{-}0.370&  \phantom{-}1.000 \\  \\[-1.7ex]\hline\hline
\end{tabular}
\label{tab:rd_kpi_corr}
\end{table*}

\begin{table*}
\caption{Covariance matrix for the parameters from the fit to $D^0 \to K^-\pi^+ \pi^-\pi^+$ data. Parameters are defined in Eqs.~(\ref{eq:sigpdf}) and ~(\ref{eq:bkgpdf}). Note that $\Gamma$ and $\Delta m_0$ are measured in $\kev$. Symmetric elements are suppressed.}
\begin{tabular}{ccccccc}
\hline\hline \\[-1.7ex]
 & $\Delta m_0$ & $\epsilon$ & $N_{bkg}$ & $N_{sig}$ & $c$ & $\Gamma$ \\[-1.7ex] \\ \hline \\[-1.7ex]
$\Delta m_0$ &    \phantom{-}2.206e-13   & &&&& \\
$\epsilon$ &      \phantom{-}2.586e-10 & \phantom{-}4.605e-05     & &&& \\
$N_{bkg}$ &    \phantom{-}3.251e-06 & \phantom{-}4.233e-01 &  \phantom{-}2.259e+04  &&&\\
$N_{sig}$ &    -3.208e-06 & -4.179e-01& -1.313e+04 & \phantom{-}1.874e+05 & &\\
$c$ &   -1.742e-09& -2.021e-04& -8.226e+00 & \phantom{-}8.095e+00 & \phantom{-}1.678e-02    &\\
$\Gamma$ &  -6.213e-14 & -8.633e-09 & -1.191e-04 & \phantom{-}1.175e-04 & \phantom{-}6.072e-08 & \phantom{-}2.289e-12\\ \\[-1.7ex] \hline\hline
\end{tabular}
\label{tab:rd_k3pi_cov}
\end{table*}

\begin{table*}
\caption{Parameter correlation coefficients for the parameters from the fit to $D^0 \to K^-\pi^+ \pi^-\pi^+$ data. Parameters are defined in Eqs.~(\ref{eq:sigpdf}) and ~(\ref{eq:bkgpdf}). Note that $\Gamma$ and $\Delta m_0$ are measured in $\kev$. Symmetric elements are suppressed.}
\begin{tabular}{ccccccc}
\hline\hline \\[-1.7ex]
 & $\Delta m_0$ & $\epsilon$ & $N_{bkg}$ & $N_{sig}$ & $c$ & $\Gamma$ \\[-1.7ex] \\ \hline \\[-1.7ex]
$\Delta m_0$ &  \phantom{-}1.000 & &&&&\\
$\epsilon$ &  \phantom{-}0.081  &\phantom{-}1.000  &&&&\\
$N_{bkg}$ & \phantom{-}0.046 & \phantom{-}0.415 &  \phantom{-}1.000& &&\\
$N_{sig}$ & -0.016 & -0.142 & -0.202 &  \phantom{-}1.000  && \\
$c$ & -0.029 &-0.230 &-0.422&  \phantom{-}0.144&  \phantom{-}1.000 &\\
$\Gamma$ & -0.087& -0.841& -0.524&  \phantom{-}0.179&  \phantom{-}0.310&  \phantom{-}1.000 \\  \\[-1.7ex] \hline\hline
\end{tabular}
\label{tab:rd_k3pi_corr}
\end{table*}


\end{document}